\documentclass[aps,prd,twocolumn,superscriptaddress,showpacs]{revtex4}

\usepackage{graphicx}
\usepackage{dcolumn}
\usepackage{bm}
\usepackage{longtable}

\begin{document}

\title {Measurement of Particle Production and Inclusive Differential Cross Sections in $p\bar p$ Collisions at $\sqrt{s}=1.96$~TeV}

\begin{abstract}
We report a set of measurements of particle production in inelastic $p\bar{p}$ collisions collected 
with a minimum-bias trigger at the Tevatron Collider with the CDF~II experiment.
The inclusive charged particle transverse momentum differential cross section is measured,
with improved precision, over a range about ten times wider than in previous measurements. 
The former modeling of the spectrum appears to be incompatible with the high particle momenta observed.
The dependence of the charged particle transverse momentum on the event particle multiplicity 
is analyzed to study the various components of hadron interactions. 
This is one of the observable variables most poorly reproduced by the available Monte Carlo generators.
A first measurement of the event transverse energy sum differential cross section is also reported.
A comparison with a {\sc pythia} prediction at the hadron level is performed.
The inclusive charged particle differential production cross section is fairly well reproduced only 
in the transverse momentum range
available from previous measurements. At higher momentum the agreement is poor.
The transverse energy sum is poorly reproduced over the whole spectrum.
The dependence of the charged particle transverse momentum on the particle multiplicity
needs the introduction of more sophisticated particle production mechanisms,
such as multiple parton interactions, in order to be better explained.
\end{abstract}


\affiliation{Institute of Physics, Academia Sinica, Taipei, Taiwan 11529, Republic of China} 
\affiliation{Argonne National Laboratory, Argonne, Illinois 60439} 
\affiliation{University of Athens, 157 71 Athens, Greece} 
\affiliation{Institut de Fisica d'Altes Energies, Universitat Autonoma de Barcelona, E-08193, Bellaterra (Barcelona), Spain} 
\affiliation{Baylor University, Waco, Texas  76798} 
\affiliation{Istituto Nazionale di Fisica Nucleare Bologna, $^y$University of Bologna, I-40127 Bologna, Italy} 
\affiliation{Brandeis University, Waltham, Massachusetts 02254} 
\affiliation{University of California, Davis, Davis, California  95616} 
\affiliation{University of California, Los Angeles, Los Angeles, California  90024} 
\affiliation{University of California, San Diego, La Jolla, California  92093} 
\affiliation{University of California, Santa Barbara, Santa Barbara, California 93106} 
\affiliation{Instituto de Fisica de Cantabria, CSIC-University of Cantabria, 39005 Santander, Spain} 
\affiliation{Carnegie Mellon University, Pittsburgh, PA  15213} 
\affiliation{Enrico Fermi Institute, University of Chicago, Chicago, Illinois 60637}
\affiliation{Comenius University, 842 48 Bratislava, Slovakia; Institute of Experimental Physics, 040 01 Kosice, Slovakia} 
\affiliation{Joint Institute for Nuclear Research, RU-141980 Dubna, Russia} 
\affiliation{Duke University, Durham, North Carolina  27708} 
\affiliation{Fermi National Accelerator Laboratory, Batavia, Illinois 60510} 
\affiliation{University of Florida, Gainesville, Florida  32611} 
\affiliation{Laboratori Nazionali di Frascati, Istituto Nazionale di Fisica Nucleare, I-00044 Frascati, Italy} 
\affiliation{University of Geneva, CH-1211 Geneva 4, Switzerland} 
\affiliation{Glasgow University, Glasgow G12 8QQ, United Kingdom} 
\affiliation{Harvard University, Cambridge, Massachusetts 02138} 
\affiliation{Division of High Energy Physics, Department of Physics, University of Helsinki and Helsinki Institute of Physics, FIN-00014, Helsinki, Finland} 
\affiliation{University of Illinois, Urbana, Illinois 61801} 
\affiliation{The Johns Hopkins University, Baltimore, Maryland 21218} 
\affiliation{Institut f\"{u}r Experimentelle Kernphysik, Universit\"{a}t Karlsruhe, 76128 Karlsruhe, Germany} 
\affiliation{Center for High Energy Physics: Kyungpook National University, Daegu 702-701, Korea; Seoul National University, Seoul 151-742, Korea; Sungkyunkwan University, Suwon 440-746, Korea; Korea Institute of Science and Technology Information, Daejeon, 305-806, Korea; Chonnam National University, Gwangju, 500-757, Korea} 
\affiliation{Ernest Orlando Lawrence Berkeley National Laboratory, Berkeley, California 94720} 
\affiliation{University of Liverpool, Liverpool L69 7ZE, United Kingdom} 
\affiliation{University College London, London WC1E 6BT, United Kingdom} 
\affiliation{Centro de Investigaciones Energeticas Medioambientales y Tecnologicas, E-28040 Madrid, Spain} 
\affiliation{Massachusetts Institute of Technology, Cambridge, Massachusetts  02139} 
\affiliation{Institute of Particle Physics: McGill University, Montr\'{e}al, Qu\'{e}bec, Canada H3A~2T8; Simon Fraser University, Burnaby, British Columbia, Canada V5A~1S6; University of Toronto, Toronto, Ontario, Canada M5S~1A7; and TRIUMF, Vancouver, British Columbia, Canada V6T~2A3} 
\affiliation{University of Michigan, Ann Arbor, Michigan 48109} 
\affiliation{Michigan State University, East Lansing, Michigan  48824}
\affiliation{Institution for Theoretical and Experimental Physics, ITEP, Moscow 117259, Russia} 
\affiliation{University of New Mexico, Albuquerque, New Mexico 87131} 
\affiliation{Northwestern University, Evanston, Illinois  60208} 
\affiliation{The Ohio State University, Columbus, Ohio  43210} 
\affiliation{Okayama University, Okayama 700-8530, Japan} 
\affiliation{Osaka City University, Osaka 588, Japan} 
\affiliation{University of Oxford, Oxford OX1 3RH, United Kingdom} 
\affiliation{Istituto Nazionale di Fisica Nucleare, Sezione di Padova-Trento, $^z$University of Padova, I-35131 Padova, Italy} 
\affiliation{LPNHE, Universite Pierre et Marie Curie/IN2P3-CNRS, UMR7585, Paris, F-75252 France} 
\affiliation{University of Pennsylvania, Philadelphia, Pennsylvania 19104}
\affiliation{Istituto Nazionale di Fisica Nucleare Pisa, $^{aa}$University of Pisa, $^{bb}$University of Siena and $^{cc}$Scuola Normale Superiore, I-56127 Pisa, Italy} 
\affiliation{University of Pittsburgh, Pittsburgh, Pennsylvania 15260} 
\affiliation{Purdue University, West Lafayette, Indiana 47907} 
\affiliation{University of Rochester, Rochester, New York 14627} 
\affiliation{The Rockefeller University, New York, New York 10021} 
\affiliation{Istituto Nazionale di Fisica Nucleare, Sezione di Roma 1, $^{dd}$Sapienza Universit\`{a} di Roma, I-00185 Roma, Italy} 

\affiliation{Rutgers University, Piscataway, New Jersey 08855} 
\affiliation{Texas A\&M University, College Station, Texas 77843} 
\affiliation{Istituto Nazionale di Fisica Nucleare Trieste/Udine, I-34100 Trieste, $^{ee}$University of Trieste/Udine, I-33100 Udine, Italy} 
\affiliation{University of Tsukuba, Tsukuba, Ibaraki 305, Japan} 
\affiliation{Tufts University, Medford, Massachusetts 02155} 
\affiliation{Waseda University, Tokyo 169, Japan} 
\affiliation{Wayne State University, Detroit, Michigan  48201} 
\affiliation{University of Wisconsin, Madison, Wisconsin 53706} 
\affiliation{Yale University, New Haven, Connecticut 06520} 
\author{T.~Aaltonen}
\affiliation{Division of High Energy Physics, Department of Physics, University of Helsinki and Helsinki Institute of Physics, FIN-00014, Helsinki, Finland}
\author{J.~Adelman}
\affiliation{Enrico Fermi Institute, University of Chicago, Chicago, Illinois 60637}
\author{T.~Akimoto}
\affiliation{University of Tsukuba, Tsukuba, Ibaraki 305, Japan}
\author{B.~\'{A}lvarez~Gonz\'{a}lez$^t$}
\affiliation{Instituto de Fisica de Cantabria, CSIC-University of Cantabria, 39005 Santander, Spain}
\author{S.~Amerio$^z$}
\affiliation{Istituto Nazionale di Fisica Nucleare, Sezione di Padova-Trento, $^z$University of Padova, I-35131 Padova, Italy} 

\author{D.~Amidei}
\affiliation{University of Michigan, Ann Arbor, Michigan 48109}
\author{A.~Anastassov}
\affiliation{Northwestern University, Evanston, Illinois  60208}
\author{A.~Annovi}
\affiliation{Laboratori Nazionali di Frascati, Istituto Nazionale di Fisica Nucleare, I-00044 Frascati, Italy}
\author{J.~Antos}
\affiliation{Comenius University, 842 48 Bratislava, Slovakia; Institute of Experimental Physics, 040 01 Kosice, Slovakia}
\author{G.~Apollinari}
\affiliation{Fermi National Accelerator Laboratory, Batavia, Illinois 60510}
% ...

\collaboration{CDF Collaboration\footnote{With visitors from $^a$University of Massachusetts Amherst, Amherst, Massachusetts 01003,
$^b$Universiteit Antwerpen, B-2610 Antwerp, Belgium, 
$^c$University of Bristol, Bristol BS8 1TL, United Kingdom,
$^d$Chinese Academy of Sciences, Beijing 100864, China, 
$^e$Istituto Nazionale di Fisica Nucleare, Sezione di Cagliari, 09042 Monserrato (Cagliari), Italy,
$^f$University of California Irvine, Irvine, CA  92697, 
$^g$University of California Santa Cruz, Santa Cruz, CA  95064, 
$^h$Cornell University, Ithaca, NY  14853, 
$^i$University of Cyprus, Nicosia CY-1678, Cyprus, 
$^j$University College Dublin, Dublin 4, Ireland,
$^k$University of Edinburgh, Edinburgh EH9 3JZ, United Kingdom, 
$^l$University of Fukui, Fukui City, Fukui Prefecture, Japan 910-0017
$^m$Kinki University, Higashi-Osaka City, Japan 577-8502
$^n$Universidad Iberoamericana, Mexico D.F., Mexico,
$^o$University of Iowa, Iowa City, IA  52242,
$^p$Queen Mary, University of London, London, E1 4NS, England,
$^q$University of Manchester, Manchester M13 9PL, England, 
$^r$Nagasaki Institute of Applied Science, Nagasaki, Japan, 
$^s$University of Notre Dame, Notre Dame, IN 46556,
$^t$University de Oviedo, E-33007 Oviedo, Spain, 
$^u$Texas Tech University, Lubbock, TX  79609, 
$^v$IFIC(CSIC-Universitat de Valencia), 46071 Valencia, Spain,
$^w$University of Virginia, Charlottesville, VA  22904,
$^x$Bergische Universit\"at Wuppertal, 42097 Wuppertal, Germany,
$^{ff}$On leave from J.~Stefan Institute, Ljubljana, Slovenia, 
}}
\noaffiliation

\maketitle

\section{\label{sec:Intro} Introduction}
In hadron collisions, hard interactions are theoretically well defined and described as collisions 
of two incoming partons along with softer interactions from the remaining partons.
The so-called ``minimum-bias'' (MB) interactions, on the contrary, 
can only be defined through a description of the experimental apparatus that triggers the collection 
of the data. Such a trigger is set up so as to collect, with uniform acceptance, events from all possible
inelastic interactions.
At the energy of the Tevatron, MB data consist largely of the softer inelastic interactions.
In this paper, only the inelastic particle production in the central part of the region
orthogonal to the beam axis is exploited. The diffractive interactions are neglected.
An exhaustive description of inelastic non-diffractive events can only be accomplished by
a non-perturbative phenomenological model such as that made available by the {\sc pythia} 
Monte Carlo generator.

The understanding of softer physics is interesting not only in its own right, but is also important 
for precision measurements of hard interactions in which soft effects need to be accounted for.
For example, an interesting discussion on how non-perturbative color reconnection effects
between the underlying event and the hard scattering partons may affect the top quark mass measurement 
can be found in~\cite{Skands}.
Also, effects due to multiple parton-parton interactions must be accounted for in MB measurements.
A detailed understanding of MB interactions is especially important in very high luminosity
environments (such as at the Large Hadron Collider) \cite{Moraes} where a large number of such interactions
is expected in the same bunch crossing.
MB physics offers a unique ground for studying both the theoretically poorly understood softer
phenomena and the interplay between the soft and the hard perturbative interactions.

The observables that are experimentally accessible in the MB final state, namely the particle inclusive
distributions and correlations, represent a complicated mixture of different physics effects
such that most models could readily be tuned to give an acceptable description of each single
observable, but not to describe simultaneously the entire set.
The {\sc pythia} Tune~A \cite{tuneA} event generator is, to our knowledge, the first model that 
comes close to describing a wide range of MB experimental distributions. 

In this paper three observables of the final state of antiproton-proton interactions 
measured with the CDF detector at $\surd{s}=1.96$~TeV  are presented:
1) the inclusive charged particle transverse momentum ($p_T$) \cite{coords} differential cross section,
2) the transverse energy sum ($\sum E_{T}$) differential cross section, and
3) the dependence of the charged particle average transverse momentum on the
charged particle multiplicity, $C_{\langle p_T\rangle \, \mathrm{vs} \, N_{ch}}$.

The first two measurements address two of the basic features of inelastic inclusive particle production.
The measurement of the event transverse energy sum is new to the field, and represents a first
attempt at describing the full final state including neutral particles. In this regard, it is 
complementary to the charged particle measurement in describing the global features of the inelastic 
$p\bar{p}$ cross section.
In this article, previous CDF measurements~\cite{ptpaper}~\cite{run1paper} are widely extended in range 
and precision.
The single particle $p_T$ spectrum now extends to over 100~GeV/$c$, and enables verification of the 
empirical modeling~\cite{modellisoft} of minimum-bias production up to the high $p_T$ production region 
spanning more  than twelve orders of magnitude in cross section.
The $C_{\langle p_T\rangle \, \mathrm{vs} \, N_{ch}}$ is one of the variables most sensitive to the 
combination of the physical effects present in MB collisions, and is also the variable most poorly 
reproduced by the available Monte Carlo generators.
Other soft production mechanisms~\cite{modelliptm}, different from a phenomenological extrapolation of 
QCD to the non-perturbative region, might show up in the high multiplicity region of 
$C_{\langle p_T\rangle \, \mathrm{vs} \, N_{ch}}$. Should 
this be the case, we might expect to observe final-state particle correlations similar to those observed 
in ion-ion collisions \cite{ions}.

A comparison with the {\sc pythia} Monte Carlo generator model \cite{pythia_new} 
is carried out for all the distributions and correlations studied.

The rest of this paper is organized as follows: Sec.~\ref{sec:CDF} describes the detector
components most relevant to this analysis. Section~\ref{sec:Data} describes the triggers and
the datasets used, including a short description of the Monte Carlo generator tuning, the
event selection and the backgrounds.
In Sec.~\ref{sec:Corrections} the methods used to correct the data for detector inefficiency 
and acceptance are discussed. Section~\ref{sec:Sys} is devoted to the discussion of the systematic 
uncertainties. In Sec.~\ref{sec:Results} the results are presented and compared to model predictions.

\section{\label{sec:CDF} The CDF detector}
CDF~II is a general purpose detector that combines precision charged particle tracking with projective 
geometry calorimeter towers. A detailed description of the detector can be found elsewhere~\cite{CDF}. 
Here we briefly describe the detector components that are relevant to this analysis: the tracking system, 
the central calorimeters, and the forward luminosity counters.

The tracking system is situated immediately outside the beam pipe and is composed of an inner set of
silicon microstrip detectors and an outer drift chamber (COT). 
The silicon detectors are located between radii of $1.5<r<29.0$~cm, and
provide precision measurements of the track's impact parameter with respect to the primary vertex.
The innermost layer (L00) \cite{l00} is single sided, and is attached directly on the beam pipe.
Five layers of double-sided silicon microstrips (SVXII) \cite{svx} cover the pseudorapidity $|\eta|\leq2$ 
region: in each layer one side is oriented at a stereo angle with respect to the beam axis to provide 
three dimensional measurements.
The ISL \cite{isl} is located outside SVXII. It consists of one layer of silicon microstrips covering the 
region $|\eta|<1$ and of two layers at $1<|\eta|<2$ where the COT coverage is incomplete or missing.
The COT \cite{cot} is a cylindrical open-cell drift chamber with 96 sense wire layers grouped into eight
alternating superlayers of stereo and axial wires. Its active volume covers $40<r<137$~cm
and $|z|<155$~cm, thus providing fiducial coverage up to 
$|\eta|$\raisebox{-0.2ex}{$\lesssim$}$1$ 
to tracks originating within $|z|\leq60$~cm.
Outside the COT, a solenoid provides a 1.4 T magnetic field that allows the particle
momenta to be computed from the trajectory curvature. The transverse momentum resolution is
$\sigma(p_{T})/p_{T}\simeq 0.1\%\cdot p_{T} / ($GeV$/c)$ for the integrated tracking system and 
$\sigma(p_{T})/p_{T}\simeq 0.2\%\cdot p_{T} / ($GeV$/c)$ for the COT tracking alone.

Located outside the solenoid, two layers of segmented sampling
calorimeters (electromagnetic~\cite{eem} and hadronic~\cite{had}) are
used to measure the energy of the particles.  In the central region,
$|\eta|<1.1$, the calorimeter elements are arranged in a projective
tower geometry of granularity $\Delta\eta \times \Delta\phi \approx
0.11 \times 15^{\circ}$.  The electromagnetic components use
lead-scintillator sampling.  A multi-wire proportional chamber (CES)
is embedded at approximately the depth of the shower maximum.  The
hadron calorimeter uses iron absorbers and scintillators.  At normal
incidence the total depth corresponds to about 18 radiation lengths in
the electromagnetic calorimeter and 4.5 interaction lengths in the
hadronic calorimeter.

The energy resolution of the electromagnetic calorimeter is 
$\sigma(E_{T})/E_{T}=14\%/\sqrt(E_{T}$(GeV))~$\oplus$~2\% 
for electromagnetic particles. It is $\sigma(E_{T})/E_{T}=75\%/\sqrt(E_{T}$(GeV))~$\oplus$~3\% 
for single pions when using both calorimeters.

Two systems of gas Cherenkov counters (CLC)~\cite{clc}, covering the forward regions $3.7<|\eta|<4.7$,
are used to measure the number of inelastic $p\bar{p}$  collisions per bunch crossing and to determine 
the luminosity.
For triggering purposes only, this analysis exploits a Time-of-Flight detector (TOF)~\cite{tof}
located between the COT and the solenoid at a mean radius of 140~cm. The TOF consists of
216 scintillator bars with photomultipliers at each end and covers roughly $|\eta|<1$.

\section{\label{sec:Data} Data Sample and Event Selection}

This analysis is based on an integrated luminosity of 506 pb$^{-1}$ collected
with the CDF~II detector between October 2002 and August 2004.  The
data were collected with a minimum-bias trigger that operates as follows.
An antiproton-proton bunch crossing, signalled by the Tevatron
radio frequency, is defined to contain at least one $p\bar{p}$  interaction if
there is a coincidence in time of signals in both forward and backward CLC
modules. This required coincidence is the start gate of the first-level
CDF trigger (Level 1) and is the so-called minimum-bias trigger.
CDF uses a three-level trigger system that selects events to be recorded
to tape at $\sim75$~Hz from the bunch crossing rate of approximately 2.5 MHz. The minimum-bias
trigger is rate limited at Level 1 in order to keep the Level 3 output at 1 Hz.
A total of about $16\times 10^{6}$ bunch crossings was recorded.

Part of the analysis also uses data collected with a high multiplicity trigger
that selects events that passed the minimum-bias trigger precondition
and in addition have a large number of primary charged particles. It functions at Level 1 
by selecting events with at least 14 hit bars in the TOF system, a hit being defined as the 
coincidence of two signals from the photomultipliers at the two ends of each bar.
At Level 3 this trigger requires at least 22 reconstructed tracks converging to the event vertex.  
The threshold of 14 TOF signals was selected as the highest compatible with a fully efficient
trigger for events with offline charged particle multiplicity $\ge22$.
The latter threshold was dictated by the statistics available in Run~I and that
expected for Run~II.
This data sample consists of about 64000 triggered events.

For transverse energy measurements, only part of the MB sample was used.
Only runs with initial instantaneous luminosity below $50\times 10^{30}$~cm$^{-2}$s$^{-1}$ have been kept
in order to reduce the effects of event pile-up in the calorimeters. 
The total number of bunch crossings accepted in this subsample is about $11\times 10^{6}$.
The average instantaneous luminosities of the two MB samples are roughly 
$17\times 10^{30}$~cm$^{-2}$s$^{-1}$  for the energy subsample and 
$20\times 10^{30}$~cm$^{-2}$s$^{-1}$ for the full sample.

An offline event selection is applied to the recorded sample of minimum-bias triggered events.
Events that contain cosmic-ray candidates, identified by the combination of
tracking and calorimeter timing, are rejected.
Only those events collected when all the detector components were working correctly are included 
in the final reduced data sample.

\subsection{\label{sec:Data:ev} Event Selection}
Primary vertices are identified by the convergence of reconstructed tracks along the $z$-axis.
All tracks with hits in at least two COT layers are accepted.
No efficiency correction is applied to the tracks used for this task. 
Vertices are classified in several quality classes: the higher the number of tracks
and their reconstruction quality (Sec. \ref{subsec:CorrectionsA}), 
the higher the class quality assigned to the vertex. 
For vertices of lowest quality (mainly vertices with one to three tracks) a requirement
that they be symmetric is added, {\it i.e.} there must be at least one track in both the 
positive and negative rapidity regions for the vertex to be accepted as primary.
In other words, the quantity $|(N^{+}-N^{-})/(N^{+}+N^{-})|$, where $N^{\pm}$ is the number of 
tracks in the positive or negative $\eta$ hemisphere, cannot equal one.

Events are accepted that contain one, and only one, primary vertex in the fiducial region 
$|z_{vtx}|\leq40$~cm centered around the nominal CDF $z=0$ position. This fiducial interval is further 
restricted to $|z_{vtx}|\leq20$~cm when measurements with the calorimeter are involved.

The event selection described contains an unavoidable
contamination due to multiple vertices when the separation between vertices is less than the vertex
resolution in the $z$-coordinate, which is about 3~cm.
A correction for this effect is discussed in Sec. \ref{sec:Results}.

\subsection{\label{sec:Data:vrt} Trigger and Vertex Acceptance}
Due to small inefficiencies in the response of the CLC detector, the minimum-bias trigger is not
100\% efficient. The efficiency has been evaluated by monitoring the trigger with several
central high transverse energy triggers, such as those containing a high $p_T$ track, a central 
high $p_T$ electron, or a central high $E_T$ jet. The results show that the trigger efficiency 
increases with the increase of some global event variables such as central 
multiplicity and central sum $E_{T}$. 

On the other hand, the total acceptance (including the efficiency) of the trigger
has been measured by comparing it with a sample of zero-bias events collected during
the same period. The zero-bias data set is collected without any trigger requirements, simply by starting the
data acquisition at the Tevatron radio-frequency signal.
The results are in agreement with previous studies \cite{lum} and indicate
that the efficiency depends on a number of variables, most of which in some way
are related to the number of tracks present in the detector:
number of beam interactions, number of tracks, instantaneous luminosity and the CLC calibration.
We parametrized the dependence on these variables so that a correction can be applied
on an event-by-event basis.

The total MB trigger acceptance increases linearly with the instantaneous luminosity.
As a function of the number of tracks, the acceptance is well represented by a typical
turn-on curve starting at about 20\% (two tracks) and reaching its plateau
with a value between 97 and 99\% for about 15 tracks.

As stated above, the present analysis includes data collected with the
high multiplicity trigger previously described. The offline selection for
these data is the same as that for the minimum-bias. Events from the high multiplicity
trigger are accepted if they have reconstructed charged track multiplicity
at Level 3 greater than or equal to 22. This value is a compromise between the desire for larger statistics 
in the multiplicity region where the cross section drops and the available trigger bandwidth.
The trigger efficiency for this multiplicity is higher than 97\%.

The primary vertex recognition efficiency for the MB data sample is evaluated in two ways:
by comparing the number of expected vertices on the basis of the instantaneous luminosity
and by using a Monte Carlo simulation with multiple $p\bar{p}$ interactions.
This efficiency was studied as a function of various event variables and found to be
roughly flat for $|z|\leq$40~cm, but strongly dependent on the number of interactions in the bunch crossing
and on  the number of tracks available for vertex clustering. Therefore the efficiency has been parametrized 
as a function of the number of tracks and of the instantaneous luminosity.

Because of their dependence on the number of tracks in the bunch crossing, a variable closely related
to the event particle multiplicity, both the trigger and the vertex efficiencies
affect not only the total cross section but also the shape of inclusive distributions.
The efficiency values are computed on an event-by-event basis,
and are common to all the distributions analyzed.

\subsection{\label{sec:Data:bkg} Backgrounds}
Diffractive events, with final-state particles mostly confined in the forward regions,
may have some activity in the central region that enters as a background in our sample.
By assuming the following indicative values $\sigma_{ci}/\sigma_{sd}/\sigma_{dd}=$44.4/10.3/7.0 mb
for the central-inelastic, single-, and double-diffractive cross sections \cite{sigmatot}, respectively,
and knowing the relative CLC acceptances, we estimate their contribution to the MB cross section
to be approximately 6\%. Roughly the same conclusion was drawn by analyzing a sample of diffractive events
generated with the {\sc pythia} simulation and passed through a MB trigger simulation.
Considering that in about half of the diffractive events no primary vertex is 
reconstructed, we estimate that diffractive production forms up to $3.4\%$ of our MB sample 
and is concentrated in the region of low charged particle multiplicity and low $\sum E_T$.

For the energy measurements, the presence of calorimeter towers with significant energy deposits 
not due to particles originating from the $p\bar{p}$ interaction was checked.
In a sample of zero-bias events, after requiring no reconstructed tracks and no signal in the CES,
about 0.002 towers per event were found above the pedestal threshold.
This number increases with the instantaneous luminosity and is attributed to
real particles crossing the calorimeter, probably scattered back from the forward calorimeters.
The resulting average energy per event was subtracted from the measurement of each event $\sum E_T$.

\subsection{\label{sec:Data:mc} The Monte Carlo Sample}
A sample of simulated Monte Carlo (MC) events about twice the size of the data was generated with 
{\sc pythia} version 6.216~\cite{pythia}, with parameters optimized for the best reproduction of 
minimum-bias interactions.
{\sc pythia} Tune~A \cite{tuneA} describes the MB interactions starting from a leading order QCD 
$2\rightarrow 2$ matrix element augmented by initial- and final-state showers and multiple parton 
interactions \cite{MPI}, folded in with CTEQ5L parton distribution functions~\cite{CTEQ} and the Lund 
string  fragmentation model~\cite{Lund}.
To model the mixture of hard and soft interactions, {\sc pythia} introduces a $\hat{p}_{T^{0}}$ 
cut off parameter \cite{pthat} that regulates the divergence of the 2-to-2  parton-parton perturbative 
cross section at low momenta. This parameter is used also to regulate the additional parton-parton 
scatterings that may occur in the same collision.
Thus, fixing the amount of multiple-parton interactions ({\it i.e.}, setting the $p_T$ cut-off) 
allows the hard 2-to-2 parton-parton scattering to be extended all the way down to $p_{T}(hard)=0$, 
without hitting a divergence. The amount of hard scattering in simulated MB events is, therefore,
related to the activity of the so-called underlying  event in the hard scattering processes.
The final state, likewise, is subject to several effects such as the treatments of the
beam remnants and color (re)connection effects.
The {\sc pythia} Tune~A results presented here are the predictions, not fits.

The MC sample used for all the efficiency and acceptance corrections was generated 
with Tune~A and $\hat{p}_{T^{0}}=1.5$~GeV/$c$. This tuning was found to give a similar
output as the default ($\hat{p}_{T^{0}}=0$) with only slightly better reproduction
of the high $p_T$ particles and a somewhat larger particle multiplicity distribution.

The definition of primary particles was to consider all particles with mean lifetime 
$\tau>0.3\times 10^{-10}$~s  produced promptly in the $p\bar{p}$ interaction, and the
decay products of those with shorter mean lifetimes. With this definition
strange hadrons are included among the primary particles, and those that
are not reconstructed are corrected for. On the other hand, their decay
products (mainly $\pi^{\pm}$ from $K^{0}_{S}$ decays) are excluded, while those from heavier
flavor hadrons are included.

A run-dependent simulation  with a realistic distribution of multiple interactions was employed.
Events were fully simulated through the detector and successively reconstructed
with the standard CDF reconstruction chain. The simulation includes the CLC detectors
used to trigger the MB sample. 

The MC sample agrees with data within 10\% for inclusive charged particle 
$p_T$ up to about 20~GeV/$c$ (see Fig.~\ref{fig:dsdpt_vs_pythia}), and $\eta$ distributions. 
A discussion on how well the MC sample reproduces the rest of the data can be found in Sec. \ref{sec:Results}.

\section{\label{sec:Corrections} Tracking and Energy Corrections}
This section describes the procedures adopted to correct the data for detector inefficiencies
and limited acceptance, and for reconstruction errors.
First, charged particle tracks are selected in such a way as to remove the main sources of background
such as secondary particles and mis-identified tracks (Sec. \ref{subsec:CorrectionsA}).
The tracking efficiency is then computed for the selected tracks, and an appropriate correction is 
applied to the data distributions (Sec. \ref{subsec:CorrectionsB}).
The measurement of $\sum E_T$ requires a careful evaluation of the calorimeter intrinsic response and
acceptance, and of other distorting effects, especially in the
lower $E_T$ range. A correction for each of these effects is described in Sec. 
\ref{subsec:CorrectionsC} and is applied to the data.

\subsection{\label{subsec:CorrectionsA}Track Selection and Acceptance}
Reconstructed tracks are accepted if they comply with a minimal set of quality
selections including a minimum number of hits, both in axial and
stereo layers of the COT. These requirements are made more stringent if no hits in the silicon 
detectors are used.

All tracks are required to originate in a fiducial region in the plane $(d_{0};\Delta z)$,
where $d_0$ is the nearest distance, projected in the transverse plane, between the track
extrapolation and the beam axis; $\Delta z$ is the distance between
the point of closest approach of the track to the $z$-axis and the
$z$-coordinate of the event vertex.
The actual region selected in the $(d_{0};\Delta z)$ plane depends on the track itself.
Tracks reconstructed including the information from silicon detectors are selected
within $d_0<0.1$~cm; those reconstructed with no information from the silicon detectors
have worse resolution in $d_0$, and are accepted if $d_0<0.5$~cm.
A similar selection criterion is used along the beam axis: $\Delta z<1$~cm for
tracks with silicon information and $\Delta z<2$~cm for the remaining tracks.
These track selection criteria are used to select primary tracks, and were determined from MC
simulation as the ones that maximize the ratio of primary to secondary particles.

As a further requirement, primary charged particles must have a transverse
momentum greater than 0.4~GeV/$c$ and pseudorapidity $|\eta|\leq 1$ in
order to optimize the efficiency and acceptance conditions. The track sample used in this
analysis is therefore very different from the one used to reconstruct the event vertex.

The number of primary charged particles in the event after the
above selection is defined as the event multiplicity $N_{ch}$.

\subsection{\label{subsec:CorrectionsB}Tracking Efficiency}
The detector acceptance and the tracker efficiency have been analyzed with the aim 
of estimating a correction to each inclusive distribution presented in the paper.
For each track, the multiplicative correction is computed using MC as

\begin{equation}
  C( p_T, \, N_{ch}) = \frac{N^{GEN}_{primary}( p_T, \, N_{ch}) \: \mathrm{in} \: |\eta|<1}
  {N^{REC}_{primary} ( p_T,  \, N_{ch})  \: \mathrm{in} \: |\eta|<1} \; ,
\label{eq:trackcorrection}
\end{equation}

\noindent where $N^{REC}_{primary}$ is the number of tracks reconstructed as primary and 
$N^{GEN}_{primary}$ the number of generated primary charged particles.
This correction factor includes the track detection and reconstruction efficiency, the correction for 
the contamination of secondary particles (particle interaction, pair creation), particle decays 
and mis-identified tracks (in MC, reconstructed tracks that do not match to a generated charged particle).

The tracking efficiency is strongly dependent on the number of tracks with a trajectory passing
close to the event vertex. 
To avoid biases due to an incorrect multiplicity distribution in the MC generator, the correction 
factor was evaluated, as a function of $p_T$, in ten different ranges of track multiplicity.

The tracking efficiency is the largest contribution to $C$. It is about 70\% at
$p_T=0.4$~GeV/$c$ and increases to about 92\% at 5~GeV/$c$, where it reaches a plateau.

The fraction of secondary and mis-identified tracks ranges between 1 and 3\% over the whole spectrum.
The final correction is roughly flat in $\eta$ and $\phi$, and shows two broad peaks
in $z$ that correspond to the edges of the silicon detector barrels.

The total correction, as defined in Eq.~\ref{eq:trackcorrection},
includes also the smearing correction for very high $p_T$ tracks, 
where the small curvature may be a source of high dispersion in the reconstructed $p_T$ value,
and introduces a significant deviation with respect to the generated $p_T$.

The measured track $p_T$ distribution is corrected by weighting each track that enters
the distribution by the correction (computed at the $p_T$ and $N_{ch}$ values corresponding to that
specific track) and by the event-related acceptances (trigger and vertex efficiency and diffractive 
event subtraction described in Sec. \ref{sec:Data:vrt} and \ref{sec:Data:bkg}).

\begin{figure}[h]
  \begin{center}
    \includegraphics[width=9.5cm,clip]{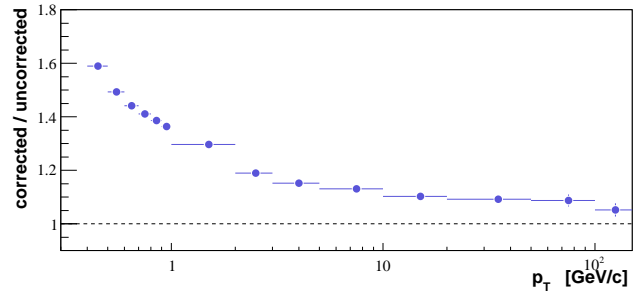}
    \caption{Ratio of the corrected to uncorrected $p_T$ distributions.
      The correction is roughly flat for $p_T>10$~GeV/$c$.}
    \label{fig:corpt}
  \end{center}
\end{figure}

To illustrate the effect of the convolution of all the corrections on the final distribution, 
the ratio of the fully corrected to the raw distributions is shown in Fig.~\ref{fig:corpt}.
The correction decreases from 1.6 at $p_T=0.4$ GeV/$c$ to 1.05 above 100 GeV/$c$.

The $C_{\langle p_T\rangle \, \mathrm{vs} \, N_{ch}}$ dependence (presented in Sec.~\ref{sec:ptave}) 
requires a specific two step correction procedure.
First, for each data point at fixed $N_{ch}$, the correction to the $\langle p_T \rangle$
is evaluated and $\langle p_T\rangle$ is corrected accordingly.
In a second step, a correction is applied for the smearing of the multiplicity of the events.
Using MC, a matrix is generated that contains the probability $P$ that an event with $n_r$
reconstructed tracks was actually generated with $n_g$ particles:

\begin{equation}
  \langle p_T \rangle_{n_{r}=m} = \sum_{i}^{n_g} (\langle p_T \rangle_{n_r=i} \cdot P_{n_r=m}^{n_g=i})  \; ,
\end{equation}

\noindent where $m$ and $i$ refer to the reconstructed and generated multiplicity bin, respectively.
In doing this it is assumed that, for all multiplicities, the average $p_T$ of events with $n_g=n$
generated tracks is the same as that of the events with $n_r=n$ reconstructed tracks.
This is indeed the case after the absolute correction on $\langle p_T \rangle$ is applied.

\subsection{\label{subsec:CorrectionsC}Calorimeter Response and Correction of the $\sum E_{T}$ Distribution}

The transverse energy is computed in the limited region 
$|\eta|<1$ as the scalar sum over the calorimeter towers of the transverse energies in the 
electromagnetic and hadronic compartments:

\begin{equation}
\sum E_{T} = \sum_{towers} E_{tower} \ \sin \left(\theta_{tower}\right)  \; ,
\end{equation}

\noindent where $\theta_{tower}$ is the polar angle measured with respect to the 
direction of the proton beam from the actual primary vertex position.
Towers with less than 100~MeV deposition are not included in the sum.

CDF calorimetry is optimized for the measurement of high energy depositions
and the analysis of its energy response is not usually performed below a few GeV.
In this paper the total $\sum E_{T}$ distribution is pushed down below this
limit and a specific study of the energy correction extension had to be done.

The calorimeter response to single charged particles was checked to be
well represented by the simulation down to a track $p_T$ of about 400~MeV/$c$.
The simulation of the energy deposition of neutral particles is assumed to be correct.
Since the fraction of charged and neutral energy produced in data and in our MC
sample agree fairly well, we rely on MC simulation to measure down to $\sum E_T =1$~GeV
the integrated calorimeter response to the total energy deposited.

\begin{figure}[ht]
  \begin{center}
    \includegraphics[width=9.5cm,clip]{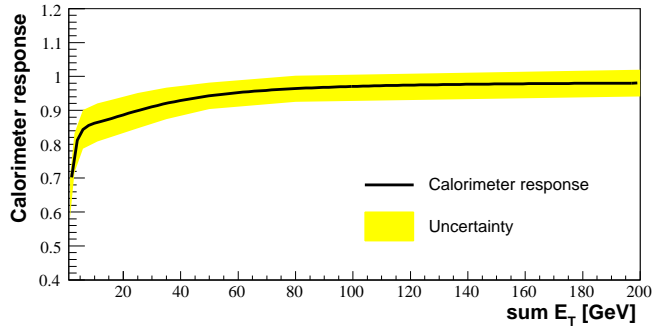}
    \caption{Calorimeter response as a function of the event $\sum E_T$.
    The systematic uncertainty is shown as a band.}
    \label{fig:sumetabscor}
  \end{center}
\end{figure}

The list of corrections applied to the data $\sum E_{T}$ distribution is the following.
All corrections are made after the calibration of the calorimeters.
\begin{itemize}
\item[1.] Tower relative correction. The response to the energy entering each calorimeter tower 
  was measured with MC as a function of the $\eta$ of the tower and of the 
  $z$ coordinate of the primary vertex
  and then normalized to the value obtained for the tower with the best response.
  This correction is introduced to make the calorimeter response flat in $\eta$ and vertex $z$.
\item[2.] Absolute correction for the calorimeter response to the total energy
  released in each event. This is calculated, using MC, as the ratio of the $\sum E_{T}$
  reconstructed in the calorimeter and corrected for the tower relative response 
  in $(\eta ;z)$, to the sum of the transverse energies of the generated primary particles in $|\eta|<1$
  whose trajectory extrapolates to the same region.
  The calorimeter response as a function of $\sum E_{T}$ is shown in Fig.~\ref{fig:sumetabscor}.
\item[3.] Correction for the different geometrical acceptance of the calorimeter 
  to events in different positions along the $z$ axis
  as a function of the $z$ coordinate of the event vertex.
  This correction ranges from 1 at $z=0$ to about 0.9 at $|z|=20$~cm.
\item[4.] Correction for undetected charged particles that curl in the magnetic field and do not
  reach the calorimeter. The average energy due to low $p_T$ charged particles,
  estimated from MC, as a function of the event $\sum E_T$, is added to each event.
\item[5.] Correction for unresolved event pile-up. Our run-dependent MC sample represents well
  the average number of multiple interactions. This was checked by plotting the ratios of the $\sum E_T$
  distributions at high luminosity to the low luminosity ones. A correction was applied by
  weighting each event by the ratio of the $\sum E_T$ distribution of the events with only 
  one generated interaction to the distribution of events with only one reconstructed interaction.
  The correction is done for five different ranges of instantaneous luminosity.
  This weight ranges from about 0.9 to about 1.1.
\item[6.] Correction for trigger and vertex acceptance and for contamination of diffractive events
  described in Sec. \ref{sec:Data:vrt} and \ref{sec:Data:bkg}, respectively.
  These corrections are applied on an event-by-event basis as weights on the $\sum E_T$
  of the events entering the final distribution.
\end{itemize}

\begin{figure}[ht]
  \begin{center}
    \includegraphics[width=9.5cm,clip]{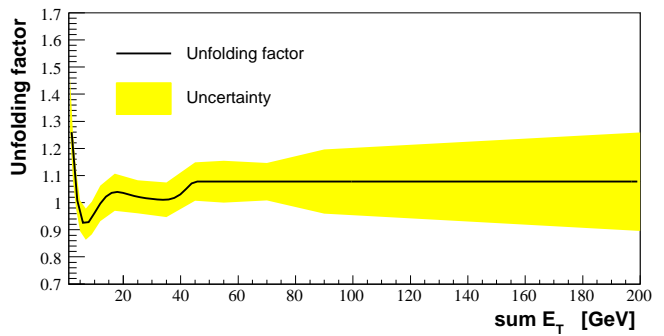}
    \caption{The unfolding factor of the $\sum E_T$ distribution. The uncertainty is taken
    as one half of the maximum variation obtained when adding and subtracting the statistical
    uncertainty to the MC distributions from which the unfolding is computed.}
    \label{fig:sumetunfold}
  \end{center}
\end{figure}
 
In terms of the calorimeter response (Fig.~\ref{fig:sumetabscor}),
the region below about 5~GeV is the most critical. The reliability of MC in evaluating the calorimeter
response was checked -- for charged particles -- against the single particle response measured from
data. A more detailed discussion can be found in Sec.~\ref{sys:mc} and leads to a systematic 
uncertainty as high as 15\% on the $\sum E_{T}$ measurement in this region.

Finally, an unfolding correction for the spread of the events with $\sum E_T$ due to the finite energy
resolution is applied. 
The unfolding is carried out in three steps. (a) An unfolding factor defined as

\begin{equation}
U(E_{T}^{gen},E_{T}^{rec})=
\frac{N_{ev}\left(\sum E_{T}^{gen}\right)}{N_{ev}\left(\sum E_{T}^{rec,corrected}\right)} \; ,
\end{equation}
where $gen$ and ${rec}$ indicate respectively the generated and the reconstructed values,
is extracted from MC; (b) in order to avoid biases due to the fact that the MC 
does not perfectly reproduce the data, {\sc pythia} Tune~A is reweighted until 
it accurately follows the data $\sum E_T$ distribution;
(c) a new unfolding factor is computed from the reweighted MC sample and is applied to the 
corrected data distribution.

The unfolding factor $U$ as a function of the event $\sum E_T$ is shown in Fig.~\ref{fig:sumetunfold}.
The final corrected $\sum E_T$ distribution is therefore obtained as

\begin{equation}
N_{ev}^{corrected}=\frac{C_5 \cdot U}{C_6} 
N_{ev}^{raw}\left(\frac{\sum E_T}{C_1 \cdot C_2 \cdot C_3}+C_4\right) \; ,
\end{equation}
where $N_{ev}^{corrected}$ and $N_{ev}^{raw}$ refer to the number of events in the corrected and raw
distributions respectively. $C_n$ refers to the $n$-th correction in the numeration given above.

\section{\label{sec:Sys}Systematic Uncertainties}

The selection criteria applied to the dataset, as well as the procedures and the MC generator
used to correct for the distortions of the apparatus, efficiency, acceptance limitation, {\it etc.}
are sources of systematic uncertainties.
Each source may affect the final distributions in different ways. A description of potential
sources of uncertainty, and the methods used to calculate their contributions
to the systematic uncertainties on the final results is presented in the following.
Table \ref{tab:syst} shows a summary of the systematic uncertainties.

\begin{table*}[t!]
\begin{center}
\begin{tabular}{lccc}
\hline
\hline
Source/Distribution$\;$ & $\;$$N_{tracks}$ ($p_T$)$\;$  & $\;$event $\langle p_T \rangle$$\;$ & $\;$$N_{events}$ ($\sum E_T$)  \\
\hline
Luminosity and Trigger &   6\%              &   ---                      &  6\%                      \\
Vertex             &  0 -- 0.6\%        &   0 -- 0.5\%               &  0 -- 2\%                 \\
Diffractive events &  0 -- 0.5\%        &   0 -- 1\%                 &  0 -- 8\%                 \\
MC tuning          &  1 -- 4 \%         &   $<1$\%                   &  5 -- 15\%                \\
Method             &  1\%               &   ---                      &  ---                      \\
Lost $E_T$         &  ---               &   ---                      & 1\%                       \\
Pile-Up            &  ---               &   ---                      & 0 -- 3\%                  \\
\hline
\hline
\end{tabular}
\caption{Summary of the systematic uncertainties.}
\label{tab:syst}
\end{center}
\end{table*}

\subsection{Integrated Luminosity, Trigger Efficiency}
There is an overall global 6\% systematic uncertainty on the effective time-integrated luminosity 
measurement \cite{Xs} that is to be added to all the cross section measurements.

Since the trigger uses the same sub-detectors as the luminosity measurement, the uncertainty on 
the trigger efficiency  is already included in the systematic uncertainty on the integrated 
luminosity measurement.

\subsection{Vertex Selection and Efficiency}
The final cross sections depend on the correction for vertex reconstruction inefficiency that was 
evaluated with MC. 
This correction, applied to the MC sample itself, returns a number
of reconstructed vertices that differs by $0.2\%$ from the number of generated ones.
The variation on the track $p_T$ distribution from this effect is minor: it has a 
maximum of $0.6\%$ at $p_{T}=1$~GeV/$c$ and is negligible above 5~GeV/$c$.
On the event $\langle p_T \rangle$ the variation is about $0.5\%$ in the multiplicity region between 1 and 5. 
On the $\sum E_T$ distribution it is larger: from $2\%$ at $E_{T}=1$~GeV 
to a negligible value above 6~GeV.

\subsection{Background of Diffractive Events}
There are two possible uncertainties on the correction for the contamination of diffractive
events: the value of the diffractive cross section with respect to the
inelastic non-diffractive  one, and the average number of diffractive particles in the COT region.
We let the contribution of diffractive events in MB vary from 5 to 7\% and the average multiplicity
from 1.0 to 1.4 tracks per event. These values are estimates of the contribution of diffractive processes 
to the inelastic central production.
We take as the uncertainty the maximum variation obtained, which is
about 30\% of the correction itself. The correction piles up in the low multiplicity region.
This uncertainty affects the track cross section by $<0.5\%$ at $p_T<1$~GeV/$c$,
the event $\langle p_T \rangle$ by less than $1\%$ in the first two multiplicity bins,
and the $\sum E_T$ cross section by 8 to 1\% in $\sum E_T<10$~GeV.

\subsection{\label{sys:mc} Uncertainties Related to the MC Generator}
The Monte Carlo modeling of any of the kinematic distributions of particles always introduces an
uncertainty on the corrections when the data distributions are not well reproduced.
To evaluate this uncertainty, a second sample of events was simulated with the
same Monte Carlo generator but different tuning (tune~DW \cite{tuneDW}). This tuning, when employed
for MB production, yields less energy per event than both data and Tune~A.

The track reconstruction has a small, but non zero, inefficiency in any kinematic variable.
The difference produced by different {\sc pythia} configurations on the final corrected
distributions is taken as a systematic uncertainty. We find that the corrected track $p_T $ distribution
varies by $1$ to $4$\% and the $C_{\langle p_T\rangle \, \mathrm{vs} \, N_{ch}}$ 
dependence varies by less than 1\%.
To avoid biases due to an incorrect multiplicity distribution in the MC generator, the correction was
evaluated in different multiplicity bins. We compare the distributions corrected inclusively 
(integrating over all particle multiplicities) and differentially with respect to the multiplicity, 
and we find a relative difference of about $1\%$ over the whole $p_T$ spectrum.

Another uncertainty is due to the contamination of secondary particles.
To address this effect, our selection (track $d_0$ and $\Delta z$) is varied both in data
and MC and the resulting average number of tracks is compared. No significant variations were
observed, after correction, on the average multiplicity.

For the energy measurement, the largest uncertainty is  due to the simulation of 
neutral particles, including the detector simulation and the particle generator. There is no way to 
disentangle these effects, but their combination may be reflected by a different fraction of neutral energy
in MC and in data. This, in turn, may affect the global correction since the energy from neutral particles
has a higher calorimeter response than the energy from charged particles.
The observed difference in neutral fraction from 0.42 to 0.48 (average values) in data, with respect 
to MC, corresponds to a variation in the calorimeter response to $\sum E_T$ by 2\%.

We take the difference between the $\sum E_T$ distributions corrected with different MC tunings
as the uncertainty due to the generator. The uncertainty is about 15\% at $E_T<5$~GeV, drops to about 
5\% at 10~GeV  and then remains roughly constant. 
Note that, at least in part, this uncertainty includes the previous one concerning the simulation
of neutral particles.

The uncertainty on the amount of energy per event  due to low $p_T$ looping charged particles
depends directly on the generator because the region of lower momenta is difficult to compare to data.
The two {\sc pythia} tunings that we employ give a difference of about 1\% in $\sum E_T$
over the whole spectrum, which corresponds to about the same uncertainty on the distribution shape.

\subsection{Uncertainties Originating from Event Pile-Up}
Finally, there is an uncertainty due to unresolved pile-up of events within 3~cm to each other
along the beam line. None of the algorithms that we tried was able to separate these overlaps efficiently.

\begin{figure}[h]
  \begin{center}
    \includegraphics[width=9cm,clip]{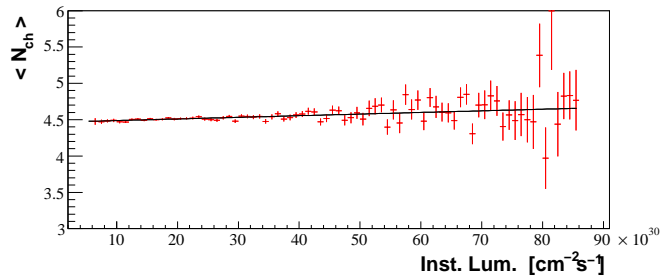}
    \caption{The raw event average charged particle multiplicity as a function of the instantaneous luminosity.
    The line represents a linear fit (with slope equal to 0.0022$\pm$0.0003). The uncertainty is 
    statistical only.}
    \label{fig:lumsys}
  \end{center}
\end{figure}

The impact on $N_{ch}$ was estimated by comparing the average multiplicity at different instantaneous
luminosities and it was found to be $<0.15$ tracks per event, this being the difference in multiplicity
between lowest and highest luminosity regions (Fig.~\ref{fig:lumsys}).
For the uncertainty  on the total number of particles in the whole MB sample, we take the difference 
in multiplicity between the lower and the average luminosity: about 0.04 tracks per event, corresponding 
to $<1\%$ of the average raw multiplicity.

The contribution from such events has been taken into account when counting the number of events
that enter the cross section calculation (Sec.~\ref{sec:ptinc}), but an uncertainty on the correction 
remains. 
It amounts to 0.005 tracks per event, which corresponds to a variation of 0.1\% of the total MB cross 
section.

The impact on the average track $p_T$ is negligible; the maximum variation
observed when varying the luminosity is about 0.004~GeV/$c$. The uncertainty on the shape of the 
distribution is therefore negligible.

The effect on $C_{\langle p_T\rangle \, \mathrm{vs} \, N_{ch}}$ is also negligible. 
This becomes clearer when considering that since the effect 
on the $p_T$ is almost zero, any variation could only be due to the reallocation of events along the 
multiplicity axis.
The ratio of two plots from samples of high and low luminosities shows negligible variation.

In the case of the energy measurement, the effect of undetected pile-up is much larger and was corrected 
for (Sec.~\ref{sec:et}), but a small uncertainty still remains on the correction itself due to the 
uncertainty on the calibration of the MC pile-up process.

We may assume that there is no pile-up below a given luminosity ({\it e.g.}, 
$10\times 10^{30}$~cm$^{-2}s^{-1}$) and use this low luminosity sample to compare to our distribution. 
The ratio of the two is compatible with unity.
However, although the pile-up probability in the low luminosity sample is small ($<1\%$), it is not 
negligible.
We may then assume an uncertainty proportional to that of the MB inelastic non-diffractive
cross section used by the MC generator.
By assuming conservatively an uncertainty of the MB inelastic non-diffractive cross section 
used by the MC generator of 6~mb,
we calculate that 
this is equivalent to a variation in the sample average luminosity of $2.5\times 10^{30}$~cm$^{-2}s^{-1}$,
which would be reflected as a $\Delta(\sum E_{T})$ of $\pm0.04$~GeV. 
This, in turn, corresponds to an uncertainty on the distribution of $<3$\% at $E_T=2$~GeV 
and negligible at $E_T>4$~GeV.

\begin{table*}[htb!]
\begin{center}
\begin{tabular}{lccccc}
\hline
\hline
      &  $p_0$           &       $n$     &  $s$          & $p_T$ range (GeV/$c$) &  $\chi^{2}/dof$  \\
\hline
Run 0, 1800~GeV (Eq.\ref{eq:fit}) & 1.29$\pm$0.02    & 8.26$\pm$0.08  & --  & 0.4 - 10.   &  102/64  \\
Run 0, 1800~GeV (Eq.\ref{eq:fit})  & 1.29$\pm$0.02    & 8.26$\pm$0.07  & --  & 0.5 - 10.   &  90/62   \\
Run 0, 1800~GeV (Eq.\ref{eq:fit})  & 1.3 fixed        & 8.28$\pm$0.02  & --  & 0.4 - 10.   &  103/65  \\
\hline
Run II, 1960~GeV (Eq.\ref{eq:fit}) & 1.230$\pm$0.004  & 8.13$\pm$0.01 & --  & 0.4 - 10.  &  352/192  \\
Run II, 1960~GeV (Eq.\ref{eq:fit})  & 1.223$\pm$0.005  & 8.11$\pm$0.01 & -- & 0.5 - 10.  &  258/182  \\
\hline
Run II, 1960~GeV (Eq.\ref{eq:fit2}) & 1.29$\pm$0.02  & 8.30$\pm$0.07  & 4.3$\pm$0.1   & 0.4 - 150. &  94/233  \\
Run II, 1960~GeV (Eq.\ref{eq:fit2}) & 1.36$\pm$0.04  & 8.47$\pm$0.09  & 4.64$\pm$0.07 & 0.5 - 150. &  80/223  \\
\hline
\hline
\end{tabular}
\caption{Comparison of fit parameters with the 1988 data (Run~0). The region $0.4<p_T<0.5$~GeV/$c$ 
in Run~0 data had a large uncertainty on the track efficiency.
The two lower rows refer to a fit with the function described in Eq.~\ref{eq:fit2}.}
\label{tab:ptfit}
\end{center}
\end{table*}

\subsection{Total Systematic Uncertainties}
All the sources  of uncertainty mentioned in Sec.\ref{sec:Sys} add up to the total systematic 
uncertainty that we attribute to each distribution as shown in the relative plots.
Those originating from MC are added linearly, and their sum is added in quadrature with the others.
Uncertainties arising due to the finite MC statistics used to calculate the corrections are represented
in the error bars on the data points; their contribution is about 50\%.
For the track $p_T$ distribution, the summed systematic uncertainties range between 3 and 6\%,
for the $C_{\langle p_T\rangle \, \mathrm{vs} \, N_{ch}}$ correlation from negligible values up to 1.5\%, 
and for the $E_T$ distribution from 5\% to 25\%. 
These numbers do not include the 6\% uncertainty on the integrated luminosity.

It is worth noting that in this paper the measurements of $p_T$ and $\sum E_T$
spectra are pushed down to very low particle energies. CDF II has limited sensitivity in these regions,
so that the correction must necessarily rely heavily on simulation.

\section{\label{sec:Results} Results}

\subsection{\label{sec:ptinc} Track $p_{T}$ Cross Section}

The single particle invariant cross section per unit phase-space element is defined as

\begin{equation}
E\frac{d^3\sigma}{dp^3} = \frac{d^3\sigma}{p_{T} dp_{T} d\phi dy} \; ,
\end{equation}
where $E$, $p$, and $y$ are the particle energy, momentum, and rapidity, respectively.
The charged particle $p_T$ distributions in bins of $\eta$ and $\phi$ have the same 
shape and mean values. Therefore the cross section factorizes in $\phi$ and $y$ 
and we may write the invariant $p_T$ differential form as

\begin{equation}
E\frac{d^3\sigma}{dp^3} = \frac{d^{3}\sigma}{p_{T} \Delta\phi \Delta y dp_{T}} =
\frac{N_{pcles}/ (\varepsilon \cdot A)}{\mathcal{L} p_{T} \Delta\phi \Delta y dp_{T}} \; ,
\end{equation}
where $N_{pcles}$ is the raw number of charged particles that is to be corrected for all efficiencies, 
$\varepsilon$, and acceptance $A$. $\mathcal{L}$ is the effective time-integrated luminosity of the sample.

\begin{figure*}[t!]
  \begin{center}
    \includegraphics[width=15cm,clip]{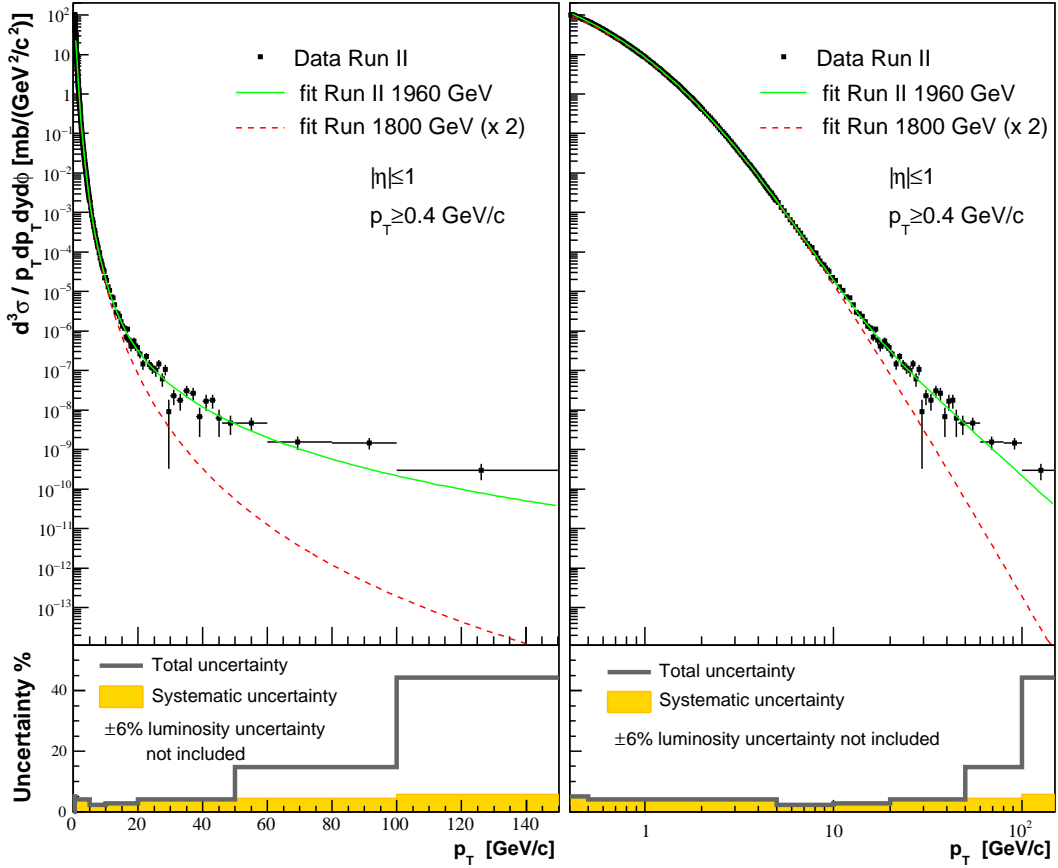}
    \caption{Left upper plot: the track $p_T$ differential cross section is shown.
    The error bars describe the uncertainty on the data points. This uncertainty includes the
    statistical uncertainty on the data and the statistical uncertainty on the total correction.
    A fit to the functional form in Eq.~\ref{eq:fit} in the region of $0.4<p_T<10$~GeV/$c$ is also
    shown for the data used in the 1988 analysis \cite{ptpaper} at the center of mass energy of
    1800~GeV (dashed line).
    A fit with a more complicated function (Eq.\ref{eq:fit2}) is shown as a continuous line.
    The fit to the 1800 GeV data is scaled by a factor 2 to account for the different 
    normalization.
    In the plot at the bottom, the systematic and the total uncertainties are shown.
    The total uncertainty is the quadratic sum of the uncertainty reported on the data points
    and the systematic uncertainty.
    The right-hand-side plots show the same distributions but with a logarithmic horizontal scale.}
    \label{fig:dsdpt}
  \end{center}
\end{figure*}

The accepted region in $\Delta y$ is calculated from the $\eta$ for each charged track,
always assuming the charged pion mass. To obtain a number of tracks
per unit rapidity interval, each track is weighted by $1/2y$ evaluated at $\eta =1$.
This procedure introduces a bias that could be avoided only by assigning the correct particle
mass to all the reconstructed tracks, which is not possible experimentally. 
Using MC, it was estimated that this bias is at most 5\% at $p_T=0.4$~GeV/$c$, and becomes negligible 
above 5~GeV/$c$. This estimate has in turn an uncertainty that is difficult to estimate due to the lack 
of measurements of the relative abundance of particles in MB data.

The acceptance $A$ takes into account the limited $z_{vertex}$ region and the rejection of crossings
with event pile-up. In the latter case the number of undetected events was
estimated indirectly  by plotting the average $N_{ch}$ as a function of the instantaneous luminosity 
(Fig.~\ref{fig:lumsys}).
In this plot, the increase in $\langle N_{ch}\rangle$ is due to the increase in number of pile-up events.
We assume that virtually no pile-up is present at a luminosity of $\mathcal{L}=1\times 10^{30}$~cm$^{-2}$s$^{-1}$.
The difference with respect to the $\langle N_{ch}\rangle$ at the average luminosity of the sample
yields the estimated number of events that went unobserved.
The final acceptance within $|\eta|<1$ of our event selections for this event sample is 
$A=0.595 \pm 0.006$.

\begin{figure*}[t!]
  \begin{center}
    \includegraphics[width=15cm,clip]{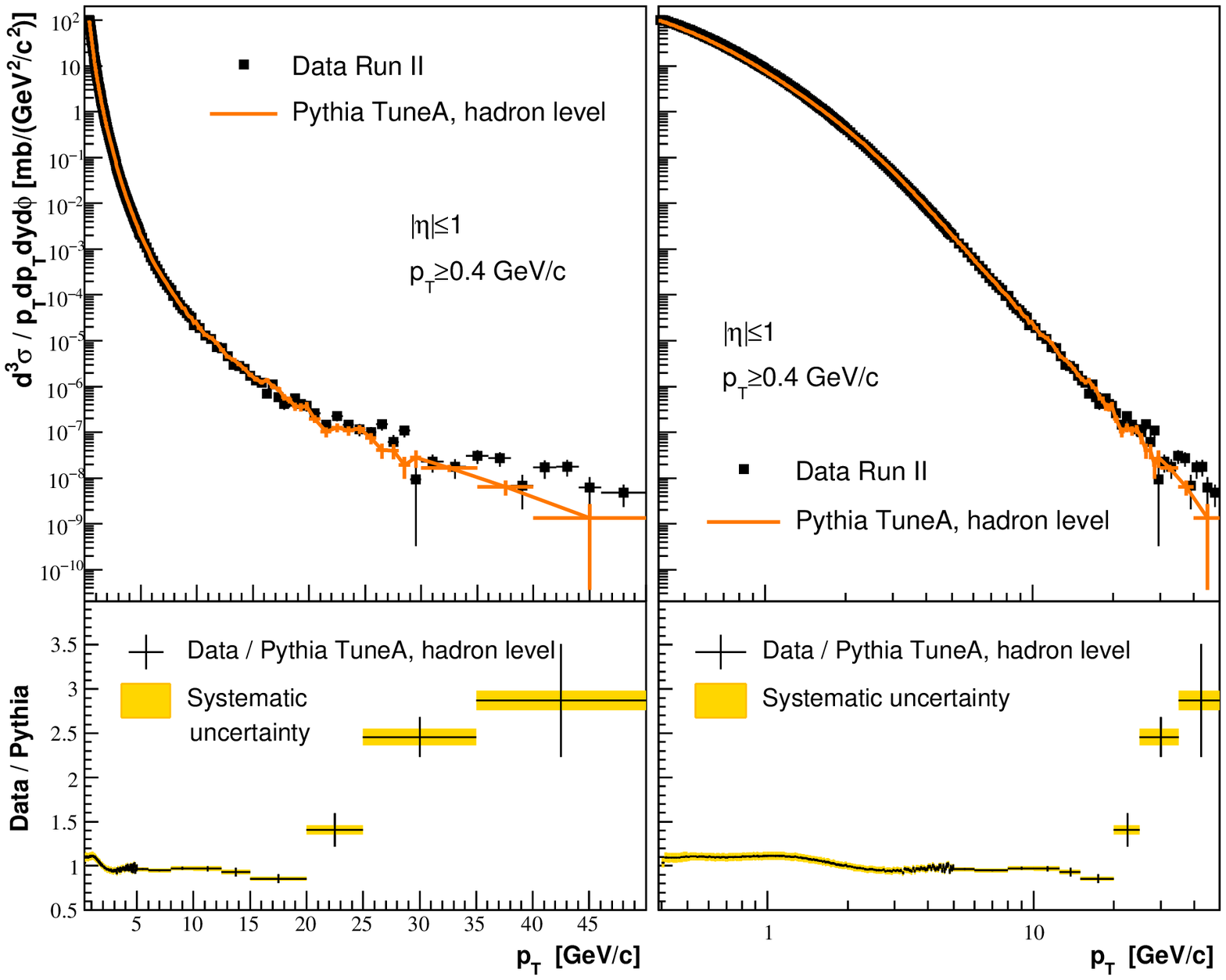}
    \caption{Left upper plot: comparison of the track $p_T$ differential cross section with 
    {\sc pythia} prediction at hadron level (Tune~A with $\hat{p}_{T^{0}}=1.5$~GeV/$c$). 
    The data error bars describe the uncertainty on the data points. This uncertainty includes the
    statistical uncertainty on the data and the statistical uncertainty on the total correction.
    The error bars on MC represent its statistical uncertainty.
    The ratio of data over prediction is shown in the lower plot. 
    The right-hand-side plots show the same distributions but with a logarithmic horizontal scale.
    Note that these distributions are cut off at 50~GeV/$c$ since {\sc pythia} does not produce particles 
    at all beyond that value.}
    \label{fig:dsdpt_vs_pythia}
  \end{center}
\end{figure*}

The differential cross section is shown in Fig.~\ref{fig:dsdpt}.
The same measurement was discussed in~\cite{studimisti}
and last published by the CDF collaboration in 1988~\cite{ptpaper}.
For historical reasons, the data published in 1988 were based on the average of positive plus 
negative tracks, {\it i.e.} only half of the total tracks were included, which explains
most of the scale factor of about 2 between the two measurements. Besides this,
the new measurement shows a cross section about 4\% higher than the previous one. 
At least part of this difference may be explained by the increased center-of-mass 
energy of the collisions from 1800 to 1960~GeV.
It should be noted, however, that in 1988 the integrated luminosity was determined indirectly
from the UA4 cross section \cite{UA4} and from the number of events selected.
In the region where the 1800 GeV data are available, the distributions have the same shape.

We observe that modeling the particle spectrum with the power-law form
used in 1988 to fit the distribution (Eq.~\ref{eq:fit}), does not account for the 
high $p_T$ tail observed in this measurement (Fig.~\ref{fig:dsdpt}).
The form in Eq.~\ref{eq:fit} is merely empirical, and the $\chi^{2}$s of the 1988 data fits
were already quite poor. Nevertheless, in the limited region up to $p_T=10$~GeV/$c$, we obtain,
for the present data, a set of fit parameters compatible with those published in 1988 
(Table~\ref{tab:ptfit}).

\begin{equation}
f = A \left(\frac{p_{0}}{p_{T}+p_{0}}\right)^{n}  \; .
\label{eq:fit}
\end{equation}

In our measurement, the tail of the distribution is at least three orders of magnitude higher than what 
could be expected by simply extrapolating to high $p_T$ the function that fits the low $p_T$ region.
In order to fit the whole spectrum, we introduced a more sophisticated parametrization (Eq.\ref{eq:fit2}):

\begin{equation}
f = A \left(\frac{p_{0}}{p_{T}+p_{0}}\right)^{n} + B\left(\frac{1}{p_T}\right)^{s} \; .
\label{eq:fit2}
\end{equation}

\noindent With this new function, we obtain a good  $\chi^{2}$ (see table~\ref{tab:ptfit}) but the 
data are still not well reproduced above about 100~GeV/$c$.

Figure~\ref{fig:dsdpt_vs_pythia} shows the ratio of data over {\sc pythia} at hadron level.
Also in this case, the data show a larger cross section at high $p_T$
starting from about 20~GeV/$c$. The MC generator does not produce any particles at all beyond 50~GeV/$c$.

\subsection{\label{sec:ptave} Mean $p_{T}$ vs Event Multiplicity}

The dependence of $p_T$ on multiplicity is computed as the average $p_T$ of all charged particles
in events with the same charged multiplicity $N_{ch}$, as a function of $N_{ch}$:

\begin{equation}
  C_{\langle p_{T}\rangle \, \mathrm{vs} \, N_{ch}}=
    \frac{\sum_{ev} \sum_{i}^{N_{ch}} p_{T}^{i}}{N_{ev}^{N_{ch}} \cdot N_{ch}} \; .
\end{equation}

The rate of change of $\langle p_T\rangle$ versus $N_{ch}$ is a measure of the amount of hard 
versus soft processes contributing to minimum-bias collisions; in simulation the rate is 
sensitive to the modeling of the multiple-parton interactions (MPI)~\cite{Skands}.
The model that currently best reproduces the correlation, {\sc pythia} Tune~A, was tuned to fit
the activity in the so-called underlying event in high transverse momentum jet production~\cite{Field2}.
However, it uses the same cut-off parameter $\hat{p}_{T^{0}}$ to regulate the divergence of the primary 
2-to-2 parton-parton scattering and the number of additional parton-parton interactions in the same 
collision. 
In addition, in {\sc pythia} the final state is subject to color (re)connection effects between different 
parton interactions of the same collision.

\begin{figure*}[p]
  \begin{center}
    \includegraphics[width=8cm,clip]{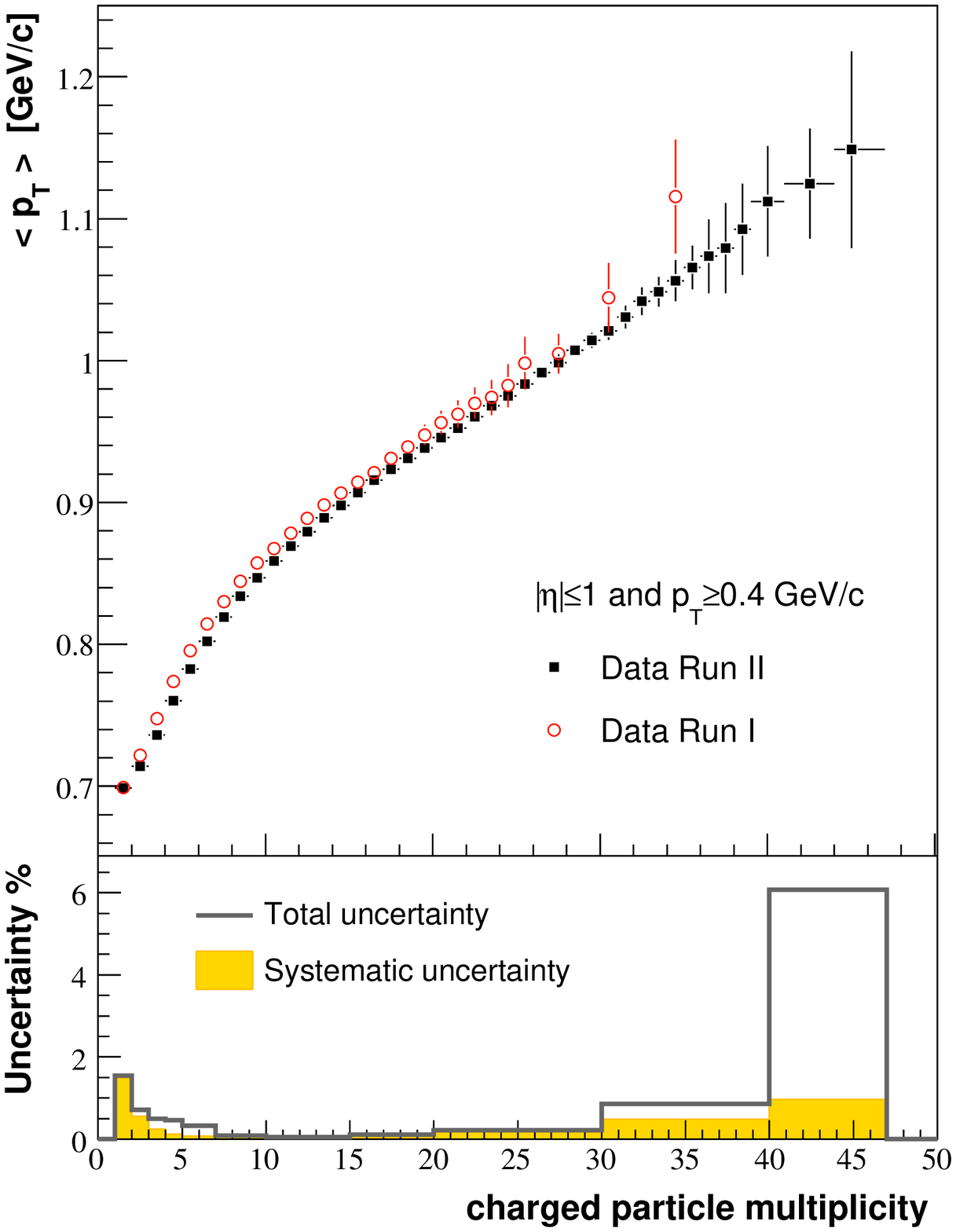}
    \caption{The dependence of the average track $p_T$ on the event multiplicity.
    A comparison with the Run~I measurement is shown.
    The error bars in the upper plot describe the uncertainty on the data points. 
    This uncertainty includes the statistical uncertainty on the data and the statistical 
    uncertainty on the total correction.
    In the lower plot the systematic uncertainty (solid yellow band) and the total
    uncertainty are shown.
    The total uncertainty is the quadratic sum of the uncertainty reported on the data points
    and the systematic uncertainty.}
    \label{fig:ptm}
    \includegraphics[width=10cm,clip]{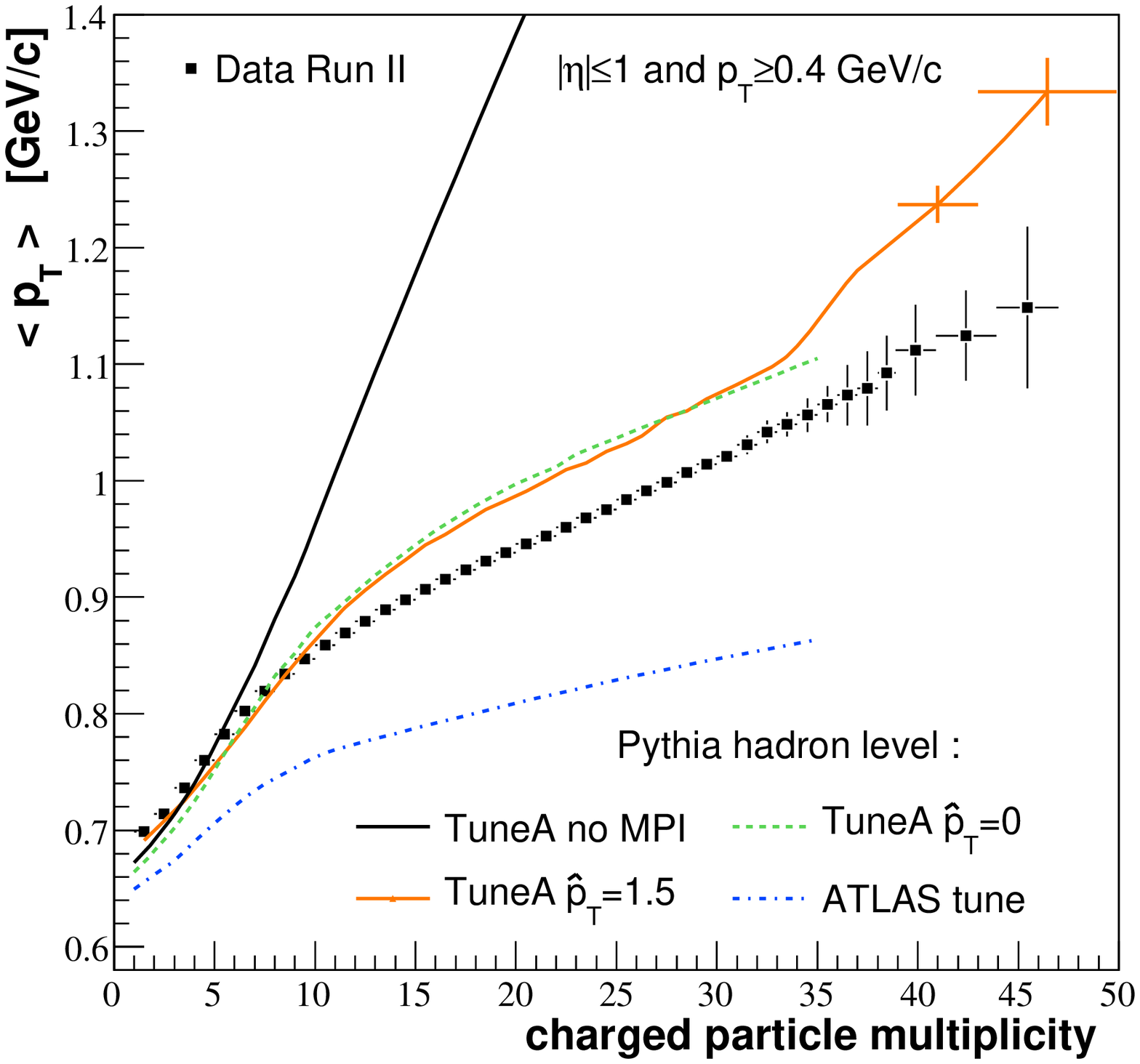}
    \caption{For tracks with $|\eta|<1$,
    the dependence of the average track $p_T$ on the event multiplicity is shown.
    The error bars on data describe the uncertainty on the data points. 
    This uncertainty includes the statistical uncertainty on the data and the statistical 
    uncertainty on the total correction.
    A comparison with various {\sc pythia} tunes at hadron level is shown. Tune~A with
    $\hat{p}_{T^{0}}=1.5$~GeV/$c$ was used to compute the MC corrections in this analysis
    (the statistical uncertainty is shown only for the highest multiplicities where it is significant).
    Tune~A with $\hat{p}_{T^{0}}=0$~GeV/$c$ is very similar to $\hat{p}_{T^{0}}=1.5$~GeV/$c$. 
    The same tuning with no multiple parton interactions allowed (``no MPI'') yields an average 
    $p_T$ much higher than data for multiplicities greater than about 5. 
    The ATLAS tune yields too low an average $p_T$ over the whole multiplicity range.}
    \label{fig:ptm2}
  \end{center}
\end{figure*}

\begin{figure*}[t!]
  \begin{center}
    \includegraphics[width=15cm,clip]{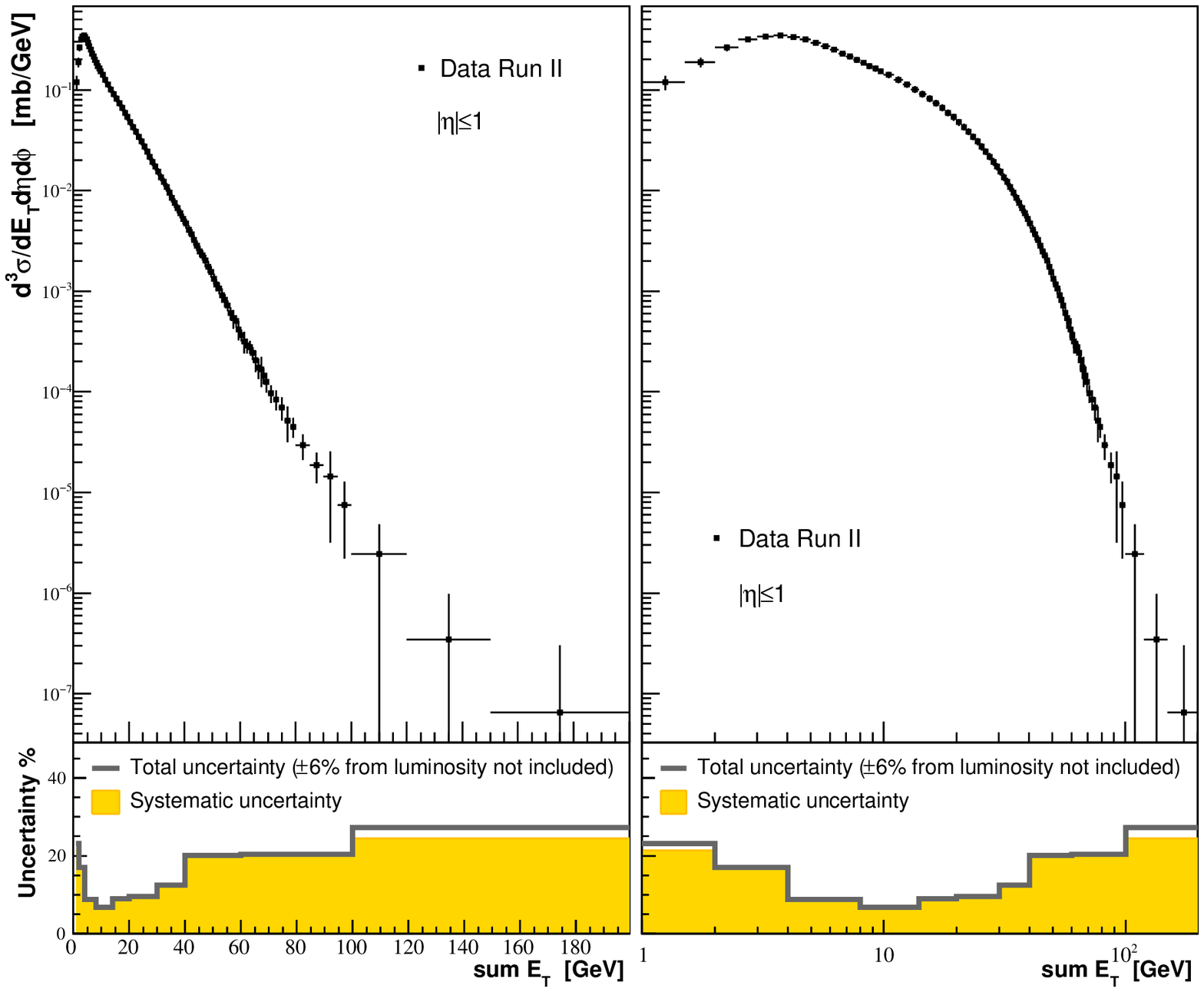}
    \caption{Left upper plot: the differential $\sum E_T$ cross section in $|\eta|\leq1$. 
    The error bars describe the uncertainty on the data points. This uncertainty includes the
    statistical uncertainty on the data and the statistical uncertainty on the total correction.
    In the bottom plot the systematic (solid band) and the total (continuous line) 
    uncertainties are shown.
    The total uncertainty is the quadratic sum of the uncertainty reported on the data points
    and the systematic uncertainty.
    The right-hand-side plots show the same distributions but with a logarithmic horizontal scale.}
    \label{fig:dsdet}
  \end{center}
\end{figure*}

The naive expectation from an uncorrelated system of strings decaying to hadrons would be 
that the $\langle p_T \rangle$ should be independent of $N_{ch}$. However, already at the ISR and 
at the $Sp\bar{p}S$ \cite{SppS}, and more recently at RHIC and at the 
Tevatron~\cite{studimisti}~\cite{studiptm}, such flat behavior was convincingly ruled out.
A study of the dependence of the mean transverse momentum $\langle p_T \rangle$ on the charged
multiplicity was already performed by CDF in Run~I and published in~\cite{run1paper}.
In the analysis presented here an extension to higher multiplicities, well over 40 particles
in the central rapidity region, is presented.
The precision greatly benefits from the larger statistics obtained with a dedicated trigger 
(Sec. \ref{sec:Data}). Data from the high multiplicity trigger are included by merging them into 
the MB sample. Comparison with Run~I data (Fig.~\ref{fig:ptm}) suggests that there is no faster rise 
of $\langle p_T \rangle$ at the higher multiplicities. Such a rise could have been considered as an indication
of a thermodynamic behavior of an expanding initial state of hadronic matter~\cite{Bjo}. 

If only two processes contribute to the MB final state, one soft, and one hard 
(the hard 2-to-2 parton-parton scattering), then demanding large $N_{ch}$ would preferentially select the
hard process and lead to a high $\langle p_T\rangle$.  However, we see from Fig.~\ref{fig:ptm2}
(Tune~A, no MPI) that with these two processes alone, the average $p_T$ increases much too rapidly.
MPI provide another mechanism for producing large multiplicities that are 
harder than the beam-beam remnants, but not as hard as the primary 2-to-2 hard scattering.
By introducing this mechanism, {\sc pythia} in the Tune~A configuration
gives a fairly good description of $C_{\langle p_T\rangle \, \mathrm{vs} \, N_{ch}}$ and, 
although the data are quantitatively not 
exactly reproduced, there is great progress over fits to Run~I data \cite{run1paper}.
Note that the systematic uncertainty is always within 2\%, a value significantly smaller
than the discrepancy with data.
{\sc pythia} Tune~A does a better job at describing the data than the ATLAS tune as described in \cite{atlas}.
Both include MPI, but with different choices for the color connections~\cite{Skands}.
In Fig.~\ref{fig:ptm2}, the ATLAS, no MPI and Tune~A $\hat{p}_{T^{0}}=0$ distributions 
do not reach multiplicities greater than about 35 solely due to the limited statistics of the 
generated samples.

\begin{figure*}[t!]
  \begin{center}
    \includegraphics[width=15cm,clip]{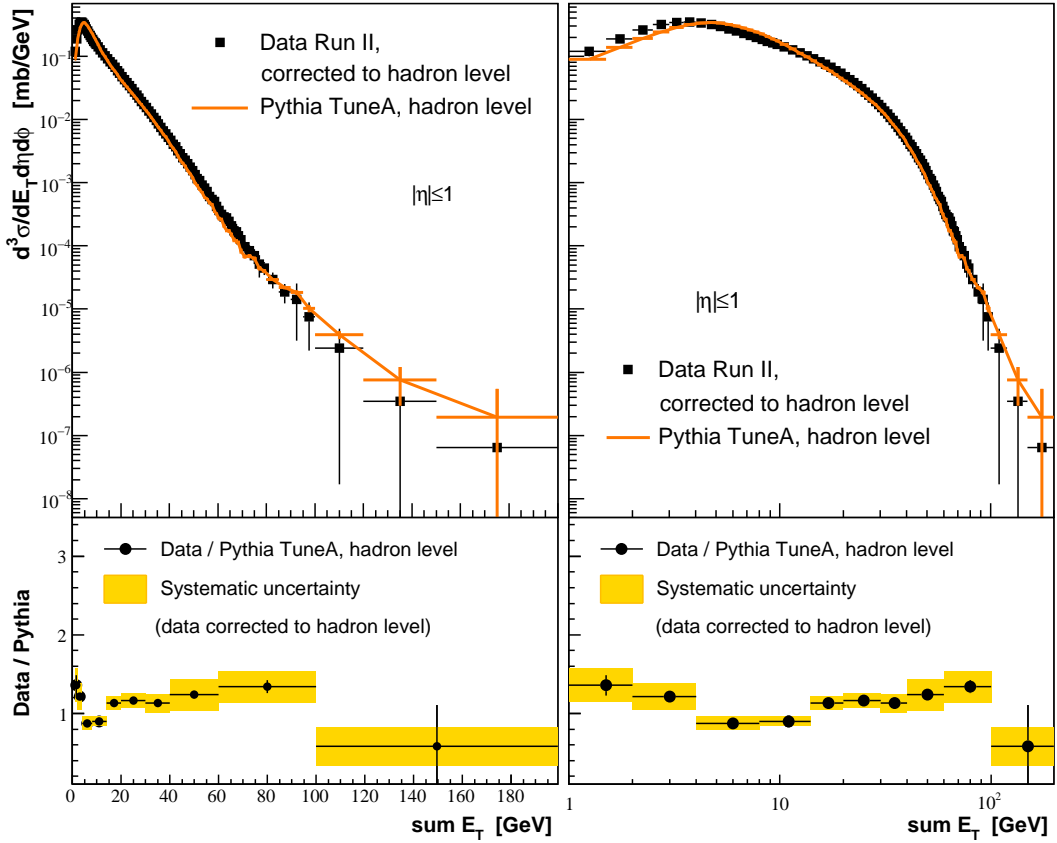}
    \caption{The left upper plot shows the same data as Fig.\ref{fig:dsdet} compared to a {\sc pythia}
      prediction at hadron level.
      The data error bars describe the uncertainty on the data points. This uncertainty includes the
      statistical uncertainty on the data and the statistical uncertainty on the total correction.
      The error bars on MC represent its statistical uncertainty.
      The ratio of data to  {\sc pythia} Tune~A using $\hat{p}_{T^{0}}=1.5$~GeV/$c$ is shown in the 
      lower plot.
      The right-hand-side plots show the same distributions but with a logarithmic horizontal scale.}
    \label{fig:dsdet_vs_pythia}
  \end{center}
\end{figure*}

\subsection{\label{sec:et} $\sum E_{T}$ Cross Section}
The $\sum E_T$ differential cross section is given by

\begin{equation}
\frac{d^3\sigma}{\Delta\phi \Delta\eta dE_{T}} =
\frac{N_{ev} / (\varepsilon \cdot A)}{\mathcal{L}\Delta\phi \Delta\eta dE_{T}} \; ,
\end{equation}

\noindent where $\mathcal{L}$
is the time-integrated luminosity for this subsample of events and $N_{ev}$ is the corresponding
corrected number of events.
The efficiency $\varepsilon$ includes all trigger and vertex efficiencies and the 
acceptance $A$ takes into account the limited $z$ region ($|z_{vtx}|<20$~cm for this analysis) 
and the rejection of crossings with event pile-up.

The differential cross section in $\sum E_T$ for $|\eta|<1$ is shown in Fig.~\ref{fig:dsdet}.
The raw and corrected event average transverse energies are $E_{T}=7.350 \pm 0.001 (\mathrm{stat.})$ 
and $E_{T}=10.4 \pm 0.2 (\mathrm{stat.}) \pm 0.7 (\mathrm{syst.})$~GeV, respectively.
This measurement, which represents the total inelastic non-diffractive cross section for events
of given $\sum E_T$, is not comparable with previous results since it is the first of its kind at
the Tevatron energies.

Figure~\ref{fig:dsdet_vs_pythia} shows a comparison with the {\sc pythia} Tune~A simulation
at hadron level. The simulation does not closely reproduce the data over the whole spectrum. 
In particular, we observe that the peak of the MC distribution is slightly shifted to higher 
energies with respect to the data.

\section{Conclusions}
Minimum-bias collisions are a mixture of hard processes (perturbative QCD) and soft processes 
(non-perturbative QCD) and, therefore, are very difficult to simulate. They contain soft 
beam-beam remnants, hard QCD 2-to-2 parton-parton scattering, and multiple parton interactions 
(soft and hard). To simulate such collisions correctly, the appropriate combination of all the 
processes involved must be known.

This paper provides a set of high precision measurements of the final state in minimum-bias interactions
and compares them to the best available MC model. The following observations may be made:
\begin{itemize}
\item[--] The former power-law modeling of the particle $p_T$ spectrum is not compatible with
the high momentum tail ($p_T$\raisebox{-0.2ex}{$\gtrsim$}$10$~GeV/$c$) 
observed in data. 
The change of slope confirms that the MB spectrum is modeled by the mixing of soft and hard
interactions. This distribution may be seen as an indirect measurement of such compositeness.
The continuity of the $p_T$ spectrum and of the $C_{\langle p_T\rangle \, \mathrm{vs} \, N_{ch}}$ 
dependence, and the absence of threshold effects
on such a large scale, indicate that there is no clear separation of hard and soft processes other
than an arbitrary experimental choice.
The more recent tunings of the {\sc pythia} MC
generator (Tune~A) reproduce the inclusive  charged particle $p_T$ distribution in data within 10\% 
up to $p_T\simeq20$~GeV/$c$ but the prediction lies below the data at high $p_T$. This may mean that the 
tune does not have exactly the right fraction of hard 2-to-2 parton-parton scattering and,
also, that there is more energy from soft processes in the data than predicted.
\item[--] The $\sum E_T$ cross section represents the first attempt to measure the neutral
particle activity in MB at CDF. The MC generator tuned to reproduce charged particle production
does not closely reproduce the shape of the distribution.
This might be related to the observation that there is an excess of energy in the underlying event 
in high transverse momentum jet production over the prediction of {\sc pythia} Tune~A.
\item[--] Among the observables in MB collisions, the dependence of the charged-particle momentum
on the event multiplicity seems to be one of the most sensitive variables to the relative
contributions by several components of MB interactions. 
This correlation is reproduced fairly well only with {\sc pythia} Tune~A: 
the mechanism of multiple parton interactions (with strong final-state correlations among them) has 
been shown to be very useful in order to reproduce high multiplicity final states with the correct 
particle transverse momenta. In fact, the data very much disfavor models without MPI, and put strong 
constraints on  multiple-parton interaction models.
\end{itemize}

The results presented here can be used to improve QCD Monte Carlo models for minimum-bias
collisions and  further our understanding of multiple parton interactions.

\section{Acknowledgments}
We thank the Fermilab staff and the technical staffs of the participating institutions for their vital 
contributions. This work was supported by the U.S. Department of Energy and National Science Foundation; 
the Italian Istituto Nazionale di Fisica Nucleare; the Ministry of Education, Culture, Sports, Science 
and Technology of Japan; the Natural Sciences and Engineering Research Council of Canada; the National 
Science Council of the Republic of China; the Swiss National Science Foundation; the A.P. Sloan Foundation; 
the Bundesministerium f\"ur Bildung und Forschung, Germany; the Korean Science and Engineering Foundation 
and the Korean Research Foundation; the Science and Technology Facilities Council and the Royal Society, UK; 
the Institut National de Physique Nucleaire et Physique des Particules/CNRS; the Russian Foundation for Basic 
Research; the Ministerio de Ciencia e Innovaci\'{o}n, and Programa Consolider-Ingenio 2010, Spain; 
the Slovak R\&D Agency; and the Academy of Finland.

\section{Appendix: Data Tables}

\renewcommand{\arraystretch}{0.7}
\renewcommand{\small}{\scriptsize}
\begin{longtable*}{ccc||ccc}
\\ \\
\caption[]{\normalsize Data of inclusive charged particle transverse momentum differential cross section
(continues across pages).}
\label{tabella1} \\
\hline\hline
$p_T$ (GeV/$c$) \ &  \ $\sigma$ (mb/(GeV$^2$/$c^2$)) \  &  \ stat. err. \  & 
\ $p_T$ (GeV/$c$)\  &  \ $\sigma$ (mb/(GeV$^2$/$c^2$)) \  &  \ stat. err. \\
\hline
\endfirsthead
\hline
$p_T$ (GeV/$c$) \ &  \ $\sigma$ (mb/(GeV$^2$/$c^2$)) \  &  \ stat. err. \  & 
\ $p_T$ (GeV/$c$)\  &  \ $\sigma$ (mb/(GeV$^2$/$c^2$)) \  &  \ stat. err. \\
\hline 
\endhead
\hline\hline
\endlastfoot
 0.40 - 0.41   &  1.0145 E+02  &  6.3 E-01 \ &   2.45 - 2.50   &  1.395 E-01  &  1.5 E-03 \\ 
 0.41 - 0.42   &  1.0215 E+02  &  6.4 E-01 \ &   2.50 - 2.55   &   1.243 E-01  &  1.4 E-03 \\ 
 0.42 - 0.43   &   9.685 E+01  &  6.2 E-01 \ &   2.55 - 2.60   &   1.111 E-01  &  1.3 E-03 \\ 
 0.43 - 0.44   &   9.245 E+01  &  6.0 E-01 \ &   2.60 - 2.65   &   1.004 E-01  &  1.1 E-03 \\ 
 0.44 - 0.45   &   8.811 E+01  &  5.8 E-01 \ &    2.65 - 2.70   &   8.97 E-02  &  1.0 E-03 \\  
 0.45 - 0.46   &   8.403 E+01  &  5.6 E-01 \ &    2.70 - 2.75   &   8.232 E-02  &  9.8 E-04 \\ 
 0.46 - 0.47   &   8.007 E+01  &  5.4 E-01 \ &    2.75 - 2.80   &   7.325 E-02  &  8.8 E-04 \\ 
 0.47 - 0.48   &   7.688 E+01  &  5.2 E-01 \ &    2.80 - 2.85   &   6.656 E-02  &  8.0 E-04 \\ 
 0.48 - 0.49   &   7.360 E+01  &  5.0 E-01 \ &    2.85 - 2.90   &   5.952 E-02  &  7.4 E-04 \\ 
 0.49 - 0.50   &   7.021 E+01  &  4.8 E-01 \ &    2.90 - 2.95   &   5.390 E-02  &  6.8 E-04 \\ 
 0.50 - 0.51   &   6.701 E+01  &  4.6 E-01 \ &    2.95 - 3.00   &   4.949 E-02  &  6.2 E-04 \\ 
 0.51 - 0.52   &   6.404 E+01  &  4.4 E-01 \ &    3.00 - 3.05   &   4.475 E-02  &  5.7 E-04 \\ 
 0.52 - 0.53   &   6.126 E+01  &  4.3 E-01 \ &    3.05 - 3.10   &   4.070 E-02  &  5.3 E-04 \\ 
 0.53 - 0.54   &   5.846 E+01  &  4.1 E-01 \ &    3.10 - 3.15   &   3.698 E-02  &  5.2 E-04 \\ 
 0.54 - 0.55   &   5.563 E+01  &  3.9 E-01 \ &    3.15 - 3.20   &   3.345 E-02  &  4.6 E-04 \\ 
 0.55 - 0.56   &   5.318 E+01  &  3.8 E-01 \ &    3.20 - 3.25   &   2.994 E-02  &  4.2 E-04 \\ 
 0.56 - 0.57   &   5.077 E+01  &  3.6 E-01 \ &    3.25 - 3.30   &   2.824 E-02  &  4.1 E-04 \\ 
 0.57 - 0.58   &   4.851 E+01  &  3.5 E-01 \ &    3.30 - 3.35   &   2.549 E-02  &  3.7 E-04 \\ 
 0.58 - 0.59   &   4.634 E+01  &  3.3 E-01 \ &    3.35 - 3.40   &   2.349 E-02  &  3.5 E-04 \\ 
 0.59 - 0.60   &   4.412 E+01  &  3.2 E-01 \ &    3.40 - 3.45   &   2.123 E-02  &  3.2 E-04 \\ 
 0.60 - 0.61   &   4.233 E+01  &  3.1 E-01 \ &    3.45 - 3.50   &   1.932 E-02  &  3.0 E-04 \\ 
 0.61 - 0.62   &   4.029 E+01  &  3.0 E-01 \ &    3.50 - 3.55   &   1.808 E-02  &  2.9 E-04 \\ 
 0.62 - 0.63   &   3.858 E+01  &  2.8 E-01 \ &    3.55 - 3.60   &   1.634 E-02  &  2.6 E-04 \\ 
 0.63 - 0.64   &   3.681 E+01  &  2.7 E-01 \ &    3.60 - 3.65   &   1.532 E-02  &  2.5 E-04 \\ 
 0.64 - 0.65   &   3.528 E+01  &  2.6 E-01 \ &    3.65 - 3.70   &   1.402 E-02  &  2.4 E-04 \\ 
 0.65 - 0.66   &   3.375 E+01  &  2.5 E-01 \ &    3.70 - 3.75   &   1.282 E-02  &  2.1 E-04 \\ 
 0.66 - 0.67   &   3.228 E+01  &  2.4 E-01 \ &    3.75 - 3.80   &   1.193 E-02  &  2.1 E-04 \\ 
 0.67 - 0.68   &   3.091 E+01  &  2.3 E-01 \ &    3.80 - 3.85   &   1.092 E-02  &  1.9 E-04 \\ 
 0.68 - 0.69   &   2.967 E+01  &  2.2 E-01 \ &    3.85 - 3.90   &   1.009 E-02  &  1.8 E-04 \\ 
 0.69 - 0.70   &   2.829 E+01  &  2.1 E-01 \ &    3.90 - 3.95   &   9.30 E-03  &  1.7 E-04 \\  
 0.70 - 0.71   &   2.715 E+01  &  2.0 E-01 \ &    3.95 - 4.00   &   8.53 E-03  &  1.6 E-04 \\  
 0.71 - 0.72   &   2.601 E+01  &  2.0 E-01 \ &    4.00 - 4.05   &   8.07 E-03  &  1.5 E-04 \\  
 0.72 - 0.73   &   2.499 E+01  &  1.9 E-01 \ &    4.05 - 4.10   &   7.46 E-03  &  1.5 E-04 \\  
 0.73 - 0.74   &   2.392 E+01  &  1.8 E-01 \ &    4.10 - 4.15   &   6.72 E-03  &  1.4 E-04 \\  
 0.74 - 0.75   &   2.293 E+01  &  1.8 E-01 \ &    4.15 - 4.20   &   6.41 E-03  &  1.3 E-04 \\  
 0.75 - 0.76   &   2.204 E+01  &  1.7 E-01 \ &    4.20 - 4.25   &   5.93 E-03  &  1.2 E-04 \\  
 0.76 - 0.77   &   2.115 E+01  &  1.6 E-01 \ &    4.25 - 4.30   &   5.39 E-03  &  1.1 E-04 \\  
 0.77 - 0.78   &   2.027 E+01  &  1.6 E-01 \ &    4.30 - 4.35   &   5.04 E-03  &  1.1 E-04 \\  
 0.78 - 0.79   &   1.943 E+01  &  1.5 E-01 \ &    4.35 - 4.40   &   4.61 E-03  &  1.0 E-04 \\  
 0.79 - 0.80   &   1.871 E+01  &  1.5 E-01 \ &    4.40 - 4.45   &   4.353 E-03  &  9.8 E-05 \\ 
 0.80 - 0.81   &   1.803 E+01  &  1.4 E-01 \ &    4.45 - 4.50   &   4.067 E-03  &  9.6 E-05 \\ 
 0.81 - 0.82   &   1.727 E+01  &  1.3 E-01 \ &    4.50 - 4.55   &   3.693 E-03  &  9.2 E-05 \\ 
 0.82 - 0.83   &   1.655 E+01  &  1.3 E-01 \ &    4.55 - 4.60   &   3.522 E-03  &  8.4 E-05 \\ 
 0.83 - 0.84   &   1.594 E+01  &  1.3 E-01 \ &    4.60 - 4.65   &   3.165 E-03  &  8.1 E-05 \\ 
 0.84 - 0.85   &   1.533 E+01  &  1.2 E-01 \ &    4.65 - 4.70   &   3.119 E-03  &  7.8 E-05 \\ 
 0.85 - 0.86   &   1.469 E+01  &  1.2 E-01 \ &    4.70 - 4.75   &   2.919 E-03  &  7.4 E-05 \\ 
 0.86 - 0.87   &   1.415 E+01  &  1.1 E-01 \ &    4.75 - 4.80   &   2.705 E-03  &  7.1 E-05 \\ 
 0.87 - 0.88   &   1.361 E+01  &  1.1 E-01 \ &    4.80 - 4.85   &   2.404 E-03  &  6.5 E-05 \\ 
 0.88 - 0.89   &   1.313 E+01  &  1.0 E-01 \ &    4.85 - 4.90   &   2.314 E-03  &  6.3 E-05 \\ 
 0.89 - 0.90   &   1.258 E+01  &  1.0 E-01 \ &    4.90 - 4.95   &   2.155 E-03  &  6.0 E-05 \\ 
 0.90 - 0.91   &   1.212 E+01  &  1.0 E-02 \ &    4.95 - 5.00   &   2.038 E-03  &  5.8 E-05 \\ 
 0.91 - 0.92   &   1.1678 E+01  &  9.6 E-02 \ &   5.00 - 5.20   &   1.784 E-03  &  3.4 E-05 \\ 
 0.92 - 0.93   &   1.1216 E+01  &  9.4 E-02 \ &   5.20 - 5.40   &   1.339 E-03  &  2.8 E-05 \\ 
 0.93 - 0.94   &   1.0829 E+01  &  9.8 E-02 \ &   5.40 - 5.60   &   1.105 E-03  &  2.3 E-05 \\ 
 0.94 - 0.95   &   1.0396 E+01  &  9.3 E-02 \ &   5.60 - 5.80   &   8.392 E-04  &  1.9 E-05 \\ 
 0.95 - 0.96   &   1.0021 E+01  &  9.1 E-02 \ &   5.80 - 6.00   &   6.59 E-04  &  1.7 E-05 \\  
 0.96 - 0.97   &   9.713 E+00  &  7.9 E-02  \ &    6.00 - 6.20   &   5.54 E-04  &  1.5 E-05 \\  
 0.97 - 0.98   &   9.325 E+00  &  7.6 E-02 \ &    6.20 - 6.40   &   4.32 E-04  &  1.2 E-05 \\  
 0.98 - 0.99   &   9.024 E+00  &  7.4 E-02 \ &    6.40 - 6.60   &   3.58 E-04  &  1.1 E-05 \\  
 0.99 - 1.00   &   8.664 E+00  &  7.0 E-02 \ &    6.60 - 6.80   &   2.979 E-04  &  9.4 E-06 \\ 
 1.00 - 1.02   &   8.227 E+00  &  6.4 E-02 \ &    6.80 - 7.00   &   2.361 E-04  &  8.4 E-06 \\ 
 1.02 - 1.04   &   7.662 E+00  &  6.0 E-02 \ &    7.00 - 7.20   &   1.999 E-04  &  7.2 E-06 \\ 
 1.04 - 1.06   &   7.129 E+00  &  5.6 E-02 \ &    7.20 - 7.40   &   1.655 E-04  &  6.4 E-06 \\ 
 1.06 - 1.08   &   6.635 E+00  &  5.3 E-02 \ &    7.40 - 7.60   &   1.422 E-04  &  5.7 E-06 \\ 
 1.08 - 1.10   &   6.188 E+00  &  4.9 E-02 \ &    7.60 - 7.80   &   1.276 E-04  &  5.3 E-06 \\ 
 1.10 - 1.12   &   5.777 E+00  &  4.6 E-02 \ &    7.80 - 8.00   &   9.60 E-05  &  4.8 E-06 \\  
 1.12 - 1.14   &   5.404 E+00  &  4.4 E-02 \ &    8.00 - 8.20   &   9.44 E-05  &  4.4 E-06 \\  
 1.14 - 1.16   &   5.057 E+00  &  4.0 E-02 \ &    8.20 - 8.40   &   7.05 E-05  &  3.7 E-06 \\  
 1.16 - 1.18   &   4.707 E+00  &  3.8 E-02 \ &    8.40 - 8.60   &   5.97 E-05  &  3.3 E-06 \\  
 1.18 - 1.20   &   4.412 E+00  &  3.6 E-02 \ &    8.60 - 8.80   &   5.02 E-05  &  3.0 E-06 \\  
 1.20 - 1.22   &   4.127 E+00  &  3.4 E-02 \ &    8.80 - 9.00   &   4.69 E-05  &  2.8 E-06 \\  
 1.22 - 1.24   &   3.858 E+00  &  3.2 E-02 \ &    9.00 - 9.20   &   3.98 E-05  &  2.6 E-06 \\  
 1.24 - 1.26   &   3.614 E+00  &  3.0 E-02 \ &    9.20 - 9.40   &   3.47 E-05  &  2.4 E-06 \\  
 1.26 - 1.28   &   3.409 E+00  &  2.8 E-02 \ &    9.40 - 9.60   &   3.23 E-05  &  2.2 E-06 \\  
 1.28 - 1.30   &   3.188 E+00  &  2.7 E-02 \ &    9.60 - 9.80   &   2.18 E-05  &  1.8 E-06 \\  
 1.30 - 1.32   &   2.985 E+00  &  2.5 E-02 \ &    9.80 - 10.00   &   2.25 E-05  &  1.8 E-06 \\ 
 1.32 - 1.34   &   2.809 E+00  &  2.4 E-02 \ &   10.00 - 10.50   &   1.89 E-05  &  1.1 E-06 \\ 
 1.34 - 1.36   &   2.6298 E+00  &  2.2 E-02 \ &  10.50 - 11.00   &   1.307 E-05  &  8.6 E-07 \\
 1.36 - 1.38   &   2.476 E+00  &  2.1 E-02 \ &   11.00 - 11.50   &   1.085 E-05  &  7.5 E-07 \\
 1.38 - 1.40   &   2.325 E+00  &  2.0 E-02 \ &   11.50 - 12.00   &   7.29 E-06  &  5.9 E-07 \\ 
 1.40 - 1.42   &   2.192 E+00  &  1.9 E-02 \ &   12.00 - 12.50   &   6.85 E-06  &  5.7 E-07 \\ 
 1.42 - 1.44   &   2.053 E+00  &  1.8 E-02 \ &   12.50 - 13.00   &   4.56 E-06  &  4.4 E-07 \\ 
 1.44 - 1.46   &   1.939 E+00  &  1.7 E-02 \ &   13.00 - 13.50   &   2.99 E-06  &  3.5 E-07 \\ 
 1.46 - 1.48   &   1.822 E+00  &  1.6 E-02 \ &   13.50 - 14.00   &   2.77 E-06  &  3.3 E-07 \\ 
 1.48 - 1.50   &   1.725 E+00  &  1.5 E-02 \ &   14.00 - 14.50   &   2.39 E-06  &  3.0 E-07 \\ 
 1.50 - 1.52   &   1.624 E+00  &  1.5 E-02 \ &   14.50 - 15.00   &   1.73 E-06  &  2.5 E-07 \\ 
 1.52 - 1.54   &   1.536 E+00  &  1.4 E-02 \ &   15.00 - 15.50   &   1.34 E-06  &  2.1 E-07 \\ 
 1.54 - 1.56   &   1.441 E+00  &  1.3 E-02 \ &   15.50 - 16.00   &   1.20 E-06  &  2.0 E-07 \\ 
 1.56 - 1.58   &   1.358 E+00  &  1.2 E-02 \ &   16.00 - 16.50   &   7.1 E-07  &  1.5 E-07 \\  
 1.58 - 1.60   &   1.287 E+00  &  1.2 E-02 \ &   16.50 - 17.00   &   1.11 E-06  &  1.8 E-07 \\ 
 1.60 - 1.62   &   1.212 E+00  &  1.1 E-02 \ &   17.00 - 17.50   &   5.9 E-07  &  1.3 E-07 \\  
 1.62 - 1.64   &   1.153 E+00  &  1.1 E-02 \ &   17.50 - 18.00   &   4.2 E-07  &  1.1 E-07 \\  
 1.64 - 1.66   &   1.084 E+00  &  1.0 E-02 \ &   18.00 - 18.50   &   4.6 E-07  &  1.1 E-07 \\  
 1.66 - 1.68   &   1.0273 E+00  &  9.7 E-03 \ &  18.50 - 19.00   &   5.5 E-07  &  1.2 E-07 \\  
 1.68 - 1.70   &   9.741 E-01  &  9.3 E-03 \ &   19.00 - 19.50   &   4.2 E-07  &  1.0 E-07 \\  
 1.70 - 1.72   &   9.176 E-01  &  8.8 E-03 \ &   19.50 - 20.00   &   3.84 E-07  &  9.8 E-08 \\ 
 1.72 - 1.74   &   8.649 E-01  &  8.3 E-03 \ &   20.00 - 21.00   &   2.61 E-07  &  5.8 E-08 \\ 
 1.74 - 1.76   &   8.238 E-01  &  8.0 E-03 \ &   21.00 - 22.00   &   1.45 E-07  &  4.1 E-08 \\ 
 1.76 - 1.78   &   7.822 E-01  &  7.6 E-03 \ &   22.00 - 23.00   &   2.27 E-07  &  5.1 E-08 \\ 
 1.78 - 1.80   &   7.389 E-01  &  7.2 E-03 \ &   23.00 - 24.00   &   1.45 E-07  &  3.9 E-08 \\ 
 1.80 - 1.82   &   6.992 E-01  &  6.9 E-03 \ &   24.00 - 25.00   &   1.16 E-07  &  3.5 E-08 \\ 
 1.82 - 1.84   &   6.612 E-01  &  6.5 E-03 \ &   25.00 - 26.00   &   1.00 E-07  &  3.1 E-08 \\ 
 1.84 - 1.86   &   6.290 E-01  &  6.3 E-03 \ &   26.00 - 27.00   &   1.48 E-07  &  3.8 E-08 \\ 
 1.86 - 1.88   &   5.963 E-01  &  6.1 E-03 \ &   27.00 - 28.00   &   6.20 E-08  &  2.4 E-08 \\ 
 1.88 - 1.90   &   5.642 E-01  &  5.7 E-03 \ &   28.00 - 29.00   &   1.08 E-07  &  3.1 E-08 \\ 
 1.90 - 1.92   &   5.382 E-01  &  5.5 E-03 \ &   29.00 - 30.00   &   9.2 E-09  &  8.9 E-09 \\  
 1.92 - 1.94   &   5.081 E-01  &  5.3 E-03 \ &   30.00 - 32.00   &   2.28 E-08  &  9.6 E-09 \\ 
 1.94 - 1.96   &   4.864 E-01  &  5.1 E-03 \ &   32.00 - 34.00   &   1.77 E-08  &  8.1 E-09 \\ 
 1.96 - 1.98   &   4.631 E-01  &  4.8 E-03 \ &   34.00 - 36.00   &   3.07 E-08  &  1.0 E-08 \\ 
 1.98 - 2.00   &   4.358 E-01  &  4.7 E-03 \ &   36.00 - 38.00   &   2.69 E-08  &  9.5 E-09 \\ 
 2.00 - 2.05   &   4.021 E-01  &  3.8 E-03 \ &   38.00 - 40.00   &   6.8 E-09  &  4.8 E-09 \\  
 2.05 - 2.10   &   3.533 E-01  &  3.3 E-03 \ &   40.00 - 42.00   &   1.69 E-08  &  7.2 E-09 \\ 
 2.10 - 2.15   &   3.125 E-01  &  3.0 E-03 \ &   42.00 - 44.00   &   1.77 E-08  &  6.9 E-09 \\ 
 2.15 - 2.20   &   2.775 E-01  &  2.7 E-03 \ &   44.00 - 46.00   &   6.2 E-09  &  4.1 E-09 \\  
 2.20 - 2.25   &   2.467 E-01  &  2.4 E-03 \ &   46.00 - 50.00   &   4.7 E-09  &  2.4 E-09 \\  
 2.25 - 2.30   &   2.194 E-01  &  2.2 E-03 \ &   50.00 - 60.00   &   4.7 E-09  &  1.7 E-09 \\  
 2.30 - 2.35   &   1.955 E-01  &  2.0 E-03 \ &   60.00 - 80.00   &   1.55 E-09  &  5.7 E-10 \\ 
 2.35 - 2.40   &   1.738 E-01  &  1.8 E-03 \ &   80.00 - 100.00   &   1.49 E-09  &  4.9 E-10 \\
 2.40 - 2.45   &   1.564 E-01  &  1.7 E-03 \ &   100.00 - 150.00   &   3.0 E-10  &  1.3 E-10 \\
\end{longtable*}
%
%
%
\vspace*{2cm}
\begin{longtable*}{ccc||ccc}
\caption{\normalsize Data of $\langle p_T \rangle$ dependence on multiplicity.}
\label{tabella2} \\
\hline
\hline
multiplicity  \  &  \  $\langle p_T \rangle$(GeV/$c$)  \ &  \ stat. err. \ &
\ multiplicity \  &  \  $\langle p_T \rangle$(GeV/$c$) \  &  \ stat. err. \\
\hline
\endfirsthead
\hline
multiplicity  &   $\langle p_T \rangle$(GeV/$c$)  &  stat. err. &
multiplicity  &   $\langle p_T \rangle$(GeV/$c$)  &  stat. err. \\
\hline
\endhead
\hline\hline
\endlastfoot
  1  &    0.6989  &   0.0016  &   22  &    0.9603  &   0.0017  \\ 
  2  &    0.7141  &   0.0016  &   23  &    0.9681  &   0.0019  \\ 
  3  &    0.7362  &   0.0016  &   24  &    0.9752  &   0.0024  \\ 
  4  &    0.7601  &   0.0017  &   25  &    0.9836  &   0.0027  \\ 
  5  &    0.7826  &   0.0018  &   26  &    0.9916  &   0.0030  \\ 
  6  &    0.8023  &   0.0018  &   27  &    0.9986  &   0.0057  \\ 
  7  &    0.8193  &   0.0019  &   28  &    1.0073  &   0.0043  \\ 
  8  &    0.8341  &   0.0019  &   29  &    1.0143  &   0.0052  \\ 
  9  &    0.8470  &   0.0003  &   30  &    1.0208  &   0.0063  \\ 
  10  &    0.8587  &   0.0004  &  31  &    1.0307  &   0.0080  \\  
  11  &    0.8694  &   0.0004  &  32  &    1.0419  &   0.0098  \\  
  12  &    0.8794  &   0.0005  &  33  &    1.049   &   0.011  \\  
  13  &    0.8891  &   0.0006  &  34  &    1.056  &   0.015  \\  
  14  &    0.8980  &   0.0006  &  35  &    1.066  &   0.015  \\  
  15  &    0.9069  &   0.0007  &  36  &    1.073  &   0.026  \\  
  16  &    0.9156  &   0.0008  &  37  &    1.079  &   0.032  \\  
  17  &    0.9235  &   0.0008  &  38  &    1.092  &   0.032  \\  
  18  &    0.9312  &   0.0010  &  39-40  &  1.112  &   0.039  \\ 
  19  &    0.9384  &   0.0011  &  41-43  &  1.125  &   0.039  \\ 
  20  &    0.9457  &   0.0013  &  44-47  &  1.149  &   0.069  \\ 
  21  &    0.9525  &   0.0015  &         &          &           \\
\end{longtable*}
%
%
%
\begin{longtable*}{ccc||ccc}
\caption{\normalsize Data of $\sum E_T$ differential cross section.}
\label{tabella3} \\
\hline
\hline
$\sum E_T$ range (GeV) \  &  \  $\sigma$ (mb/GeV) \  &  \ stat. err.\  &
\ $\sum E_T$ range (GeV) \  & \   $\sigma$ (mb/GeV) \  &  \ stat. err. \\
\hline
\endfirsthead
\hline
$\sum E_T$ range (GeV) \  &  \  $\sigma$ (mb/GeV) \  &  \ stat. err.\  &
\ $\sum E_T$ range (GeV) \  & \   $\sigma$ (mb/GeV) \  &  \ stat. err. \\
\hline
\endhead
\hline
\endfoot
\hline\hline
\endlastfoot
1.0 - 1.5  &  1.19e-01 & 1.9e-02 & 37 - 38  &  6.66e-03 & 6.7e-04 \\ 
1.5 - 2.0  &  1.89e-01 & 2.1e-02 & 38 - 39  &  5.97e-03 & 6.7e-04 \\ 
2.0 - 2.5  &  2.63e-01 & 2.0e-02 & 39 - 40  &  5.24e-03 & 6.2e-04 \\ 
2.5 - 3.0  &  3.16e-01 & 1.6e-02 & 40 - 41  &  4.72e-03 & 5.4e-04 \\ 
3.0 - 3.5  &  3.41e-01 & 1.2e-02 & 41 - 42  &  4.06e-03 & 4.6e-04 \\ 
3.5 - 4.0  &  3.46e-01 & 1.0e-02 & 42 - 43  &  3.67e-03 & 4.8e-04 \\ 
4.0 - 4.5  &  3.36e-01 & 1.0e-02 & 43 - 44  &  3.20e-03 & 3.8e-04 \\ 
4.5 - 5.0  &  3.17e-01 & 1.1e-02 & 44 - 45  &  2.84e-03 & 3.3e-04 \\ 
5.0 - 5.5  &  2.94e-01 & 1.2e-02 & 45 - 46  &  2.50e-03 & 3.3e-04 \\ 
5.5 - 6.0  &  2.72e-01 & 1.2e-02 & 46 - 47  &  2.27e-03 & 2.6e-04 \\ 
6.0 - 6.5  &  2.50e-01 & 1.2e-02 & 47 - 48  &  2.01e-03 & 2.8e-04 \\ 
6.5 - 7.0  &  2.31e-01 & 1.2e-02 & 48 - 49  &  1.75e-03 & 2.4e-04 \\ 
7.0 - 7.5  &  2.14e-01 & 1.2e-02 & 49 - 50  &  1.56e-03 & 2.0e-04 \\ 
7.5 - 8.0  &  1.99e-01 & 1.2e-02 & 50 - 51  &  1.33e-03 & 1.6e-04 \\ 
8.0 - 8.5  &  1.85e-01 & 1.2e-02 & 51 - 52  &  1.16e-03 & 1.6e-04 \\ 
8.5 - 9.0  &  1.73e-01 & 1.1e-02 & 52 - 53  &  1.07e-03 & 1.4e-04 \\ 
9.0 - 9.5  &  1.63e-01 & 1.1e-02 & 53 - 54  &  9.1e-04 & 1.5e-04 \\  
9.5 - 10  &  1.54e-01 & 1.0e-02 &  54 - 55  &  8.2e-04 & 1.2e-04 \\  
10 - 11  &  1.41e-01  & 1.0e-02 &  55 - 56  &  7.21e-04 & 8.8e-05 \\ 
11 - 12  &  1.261e-01 &  9.1e-03 & 56 - 57  &  6.05e-04 & 9.1e-05 \\ 
12 - 13  &  1.131e-01 &  8.3e-03 & 57 - 58  &  5.38e-04 & 1.1e-04 \\ 
13 - 14  &  1.013e-01 & 7.5e-03 &  58 - 59  &  5.04e-04 & 7.1e-05 \\ 
14 - 15  &  9.12e-02 & 6.8e-03 &   59 - 60  &  4.17e-04 & 9.1e-05 \\ 
15 - 16  &  8.21e-02 & 6.2e-03 &   60 - 61  &  3.67e-04 & 4.9e-05 \\ 
16 - 17  &  7.41e-02 & 5.6e-03 &   61 - 62  &  3.17e-04 & 7.7e-05 \\ 
17 - 18  &  6.66e-02 & 5.9e-03 &   62 - 63  &  2.90e-04 & 4.8e-05 \\ 
18 - 19  &  5.93e-02 & 4.5e-03 &   63 - 64  &  2.78e-04 & 4.2e-05 \\ 
19 - 20  &  5.36e-02 & 4.1e-03 &   64 - 65  &  2.43e-04 & 3.6e-05 \\ 
20 - 21  &  4.77e-02 & 3.8e-03 &   65 - 66  &  2.06e-04 & 4.9e-05 \\ 
21 - 22  &  4.28e-02 & 3.3e-03 &   66 - 67  &  1.73e-04 & 4.0e-05 \\ 
22 - 23  &  3.85e-02 & 3.1e-03 &   67 - 68  &  1.66e-04 & 5.5e-05 \\ 
23 - 24  &  3.43e-02 & 2.7e-03 &   68 - 69  &  1.45e-04 & 2.6e-05 \\ 
24 - 25  &  3.06e-02 & 2.6e-03 &   69 - 70  &  1.26e-04 & 2.8e-05 \\ 
25 - 26  &  2.72e-02 & 2.2e-03 &   70 - 72  &  9.7e-05 & 1.9e-05 \\  
26 - 27  &  2.44e-02 & 2.0e-03 &   72 - 74  &  8.4e-05 & 1.9e-05 \\  
27 - 28  &  2.18e-02 & 1.9e-03 &   74 - 76  &  7.0e-05 & 1.8e-05 \\  
28 - 29  &  1.94e-02 & 1.7e-03 &   76 - 78  &  5.2e-05 & 2.0e-05 \\  
29 - 30  &  1.73e-02 & 1.5e-03 &   78 - 80  &  4.5e-05 & 1.0e-05 \\  
30 - 31  &  1.54e-02 & 1.4e-03 &   80 - 85  &  2.9e-05 & 8.4e-06 \\  
31 - 32  &  1.35e-02 & 1.2e-03 &   85 - 90  &  1.9e-05 & 6.3e-06 \\  
32 - 33  &  1.21e-02 & 1.2e-03 &   90 - 95  &  1.4e-05 & 1.1e-05 \\  
33 - 34  &  1.08e-02 & 1.0e-03 &   95 - 100  &  7.5e-06 & 5.3e-06 \\ 
34 - 35  &  9.58e-03 & 9.7e-04 &   100 - 120  &  2.4e-06 & 2.4e-06 \\
35 - 36  &  8.49e-03 & 8.3e-04 &   120 - 150  &  3.5e-07 & 6.3e-07 \\
36 - 37  &  7.53e-03 & 8.1e-04 &   150 - 200  &  6.5e-08 & 2.4e-07 \\
\end{longtable*}
\renewcommand{\small}{\normalsize}


\begin{thebibliography}{99}
\bibitem{coords} In the CDF coordinate system, the origin is the center of the detector and the $z$-axis 
is along the beam axis. In detector coordinates, $\phi$ is defined to be the azimuthal angle around the
beam axis and $\theta$ the polar angle. The radial distance to the beam axis is referred to as $r$.
The pseudorapidity $\eta$ is defined as $\eta=-\ln\left[\tan\left(\theta/2\right)\right]$.
The transverse energy and momenta are defined by $E_{T}=E\sin\theta$ and $p_{T}=p\sin\theta$ where
$E$ is the particle's energy and $p$ the magnitude of its momentum.
\bibitem{pthat} $\hat{p}_{T^{0}}$ is defined as the transverse momentum of the 2-to-2
               scattering in the rest frame of the hard interaction.
\end{thebibliography}
\end{document}